\magnification1200

\vskip 2cm
\centerline
{\bf    Gauge fields and infinite chains of dualities }
\vskip 1cm
\centerline{ Nicolas Boulanger ${}^\star$, Per Sundell ${}^\dagger$ and Peter West ${}^\ddagger$ }
\vskip1cm
\centerline{${}^\star$ Service de M\'ecanique et Gravitation }
\centerline{Universit\'e de Mons -- UMONS}
\centerline{20 place du Parc, B-7000 Mons, Belgium}
\vskip0.5cm
\centerline{${}^\dagger$ Departamento de Ciencias F\'isicas}
\centerline{Universidad Andres Bello -- UNAB}
\centerline{Av. Rep\'ublica 252, Santiago, Chile}
\vskip0.4cm
\centerline{and }
\vskip0.4cm
\centerline{${}^\ddagger$ Department of Mathematics}
\centerline{King's College, London WC2R 2LS, UK}
\vskip 2cm

\leftline{\sl Abstract}
\bigskip
We   show that the particle states of Maxwell's theory, in $D$ dimensions,  can be represented  in an infinite number of ways by using different gauge fields. Using this result we  formulate the dynamics in terms of an 
infinite set of duality relations  which are first order in space-time derivatives.   We  derive a similar  result for the three form in eleven dimensions where such a possibility was first observed in 
the context of $E_{11}$.   We  also give an action formulation for some of the gauge fields. 
 In this paper we  give a pedagogical account of the Lorentz and gauge covariant formulation of the irreducible representations of the Poincar\'e group, used previously in higher spin  theories, as this plays a  key role in our constructions. It is clear that our results can be generalised to any particle.

\vskip2cm
\noindent

\vskip .5cm

\vfill
\eject

\medskip
\medskip
\medskip
{\bf {1. Introduction}}
\medskip
\medskip
Dirac's wish to treat the  electric and magnetic fields of Maxwell's equation in  a more  
symmetric fashion lead him to  propose the existence  of magnetic monopoles [1]. 
These were found to occur as regular solutions of spontaneously broken Yang-Mills theories 
coupled to scalar fields [2]. However, it was Montonen and Olive who proposed 
that there might be an electromagnetic duality symmetry [3] 
which was subsequently found to be present in the maximally rigid supersymmetric  theory 
. 
One of the most important discoveries in supergravity theories was the existence of 
exceptional symmetries in the maximally supergravity theories [4]. 
The scalar fields in 
these theories belong to a coset space  constructed from the exceptional symmetry. 
However, these symmetries generically act on the other fields in the supergravity theory 
and when acting on the ``spin one" fields they act as  a kind of electro-magnetic duality 
symmetry [5]. 
\par
It has been conjectured that the underlying theory of strings and branes possess a very 
large Kac--Moody symmetry called $E_{11}$ [6]. This is encoded in a non-linear realisation 
which possesses an infinite number of fields. The fields are ordered by a level and at low 
levels the fields in $E_{11}$ are just those of maximal supergravity theory in the dimension 
being considered. However the $E_{11}$ theory is democratic in that it also contains the 
dual fields as well as the traditional fields; for example, in eleven dimensions in addition to  
the graviton and three form, it contains  the six form and  a  field with the index structure 
$h_{a_1\ldots a_8, b}$ which was the dual of the gravity [6]. Indeed, it was in this paper that an equation 
of motion in $D$ dimensions  of  a field with the index structure $h_{a_1\ldots a_{D-3}, b}$ 
and the usual graviton was given at the linearised level. As this equation 
was derived form the usual formulation of gravity it was guaranteed to describe gravity in the correct way 
including the correct degrees of freedom. In fact the equation of motion of the dual graviton had been  
previously given in five dimensions in reference [7] where it was also pointed out that it had the correct 
degree of freedom to be gravity. The form of the dual graviton field had  also  previously been suggested 
in [8]. 
Finally, in [9] the action and complete set of gauge symmetries for the dual graviton in arbitrary dimension 
was constructed along the lines of [6], thereby tying 
together the results of [6] and [7]. 
\par
Although, the fields in the $E_{11}$ non-linear realisation are listed for low levels, see for example [10],  they are not systematically known at  higher levels  theory. 
However, certain results are known, these include  all the  $p$-form fields 
[11,12]  in the different dimensions, 
some of which play a key role in gauged supergravities. Also known 
are all fields that do not have blocks of ten and eleven indices in eleven dimensions [13]. 
These fields have a particularly simple form, they are just  the usual fields of the maximal 
supergravity theory,  as well as the dual fields just discussed above, as well as  an infinite 
number of fields that consist of adding blocks of $9$ indices to these fields. In eleven 
dimensions these are the fields 
$$
\{ \,h_{[1,1]}, \ A_{[3]}, \ A_{[6]}, \ h_{[8,1]},   \ A_{[9,3]}, \ A_{[9,6]}, \ 
h_{[9,8,1]},\  A_{[9,9,3]}, \ A_{[9,9,6]}, \ h_{[9, 9,8,1]}, \ldots \} 
\eqno(1.1) 
$$
where the numbers  in square brackets  indicate the number of antisymmetrised indices 
in each block.  
It was noted that when decomposed to the little group SO(9) these blocks of nine indices did 
not transform and so these fields should be just alternative ways of describing the degrees 
of freedom given in terms of the three form and graviton. 
In the sector of the graviton, this conjecture was verified in [14] at the action level. 
As a result $E_{11}$ encodes an infinite duality symmetry which should be expressed 
through an infinite series of duality 
relations which determine  the dynamics of the particle. The analogous results for other 
very extended algebras were given in [15]. 
\par 
In this paper we will show that alternative gauge field representations arise quite generally 
for any particle. Indeed, they arise naturally from the irreducible unitary representations of 
 the Poincar\'e group  $ISO(1,D-1)$ that describe any particle moving in Minkowski space-time.  
These are constructed following the method of Wigner, which involves the induced 
representation based on  the isotropy, or little,  group that preserves the momentum in a 
chosen Lorentz frame [16,17]. 
In the massless case, this method works with the gauge field 
and as a result it has a number of ad hoc steps associated with the gauge 
 transformation of this field. 
\par
However, there is another type of representation of $ISO(1,D-1)$ also capable of carrying 
the massless, irreducible and unitary representations of $ISO(1,D-1)$, and that is manifestly Lorentz 
covariant and also gauge invariant [18,19], see also [20,21,22].
This works with the fields strengths and their derivatives rather than the gauge fields. 
We note also the anterior and different method [23] where field equations for arbitrary gauge 
fields are also formulated in terms of 
curvature tensors. 
When viewed in this way we will show that there is an infinite number of ways of 
introducing different gauge potentials corresponding to the particular equations one takes 
to be Bianchi identities and those one considers to be equations of motion. This choice 
reflects the possibility of the different possible duality transformations one can carry out on 
the field strengths and all their space-time derivatives. 
In the case of the spin-2 gauge field, this mechanism was explained 
in reference [14] using the fields contained in the $E_{11}$ non-linear realisation~[6]. 
\par
In this paper we will carry out this programme for Maxwell's theory in $D$ dimensions and then 
for the three form in eleven dimensions. As is very well known the latter occurs in the eleven 
dimensional  supergravity theory [4]. 
We will find the  equations of motion for the particles when described in terms of any of the 
possible gauge fields.  We will also find an infinite series of duality relations that encode the 
dynamics of the particles and involve all the gauge potentials.  
\par
For the first sections of the paper we use familiar conventions for writing indices on fields, but as the paper 
progresses, and the number of different types of indices increases, we use a number of shorthand conventions. 
We define these as we use them, but for easy reference we give an appendix where these conventions are 
listed. 
\medskip 
\medskip 
\medskip 
{\bf 2.  Spin one and its gauge fields }
\medskip
\medskip 
In this section we illustrate the ideas of this paper in the context of the simplest model, that is, 
the Maxwell theory in $D$ dimensions. We will show that this system possesses an infinite 
number of   descriptions corresponding to an infinite number of different possible choices of  
gauge fields.  The states of any particle are the irreducible unitary representation of 
the Poincar\'e group $ISO(1,D-1)$ in $D$ dimensions. 
As we mentioned above, these were first found by Wigner [16] 
who constructed them as  an 
induced representation of $ISO(1,D-1)$ with respect to an isotropy subgroup that preserves a 
fixed momentum and it is this formulation that is most widely known. 
The representations are labelled by the representations of the isotropy group that they carry. 
In the massless helicity cases the isotropy subgroup is $SO(D-2)$ and by spin one we mean it 
carries the vector representation of $SO(D-2)$. 
\par
However, for the massless case the Wigner formulation of particle states involves 
introducing  a 
gauge field and the procedure has a number of ad hoc steps associated with the gauge 
symmetry of this field. There does exist a much less well known, but  equivalent 
formulation of the Wigner unitary irreducible representations, that is manifestly $SO(1,D-1)$ 
covariant and, for the massless case,  is also manifestly gauge invariant; 
indeed it involves field strengths and their derivatives and plays 
an important role in the formulation of nonlinear higher spin theories, see 
[19,21,22] and refs. therein.
We will ask what possible gauge 
potentials are contained in this representation and in this way we will find an infinite possible 
choices of gauge potentials.  In this section we take the opportunity to give a hopefully very 
readily understandable  account of this formulation of the irreducible representations of 
$ISO(1,D-1)$ using only knowledge that every physicists knows.

\medskip 
{\bf 2.1 The Wigner unitary irreducible representation of spin one}
\medskip

As every theorist knows a spin one particle can be described by a rank two field strength $F_{a_1a_2}$ 
subject to the Bianchi identity 
$$
\partial_{[a_1} F_{a_2a_3]}=0 
\eqno(2.1.1)$$
 and the equation of motion 
 $$
  \partial^{a} F_{ab}=0 \;.
\eqno(2.1.2)$$
These imply that 
$$
\partial^{a_1}  \partial_{[a_1} F_{a_2a_3]}=0 \quad  {\rm and \ so } 
\quad \partial^b\partial_b F_{a_1a_2}=0 \;.
\eqno(2.1.3)$$
We will hence forth denote $ F^{(0)}{}_{a_1a_2} := F_{a_1a_2}$ as this will be  the first in  
a series of objects $F^{(n)}{}_{a_1a_2 || b_{1}\ldots b_{n}} $ that we will define. 
We refer to $n$ as the level. 
\par 
To show that these do indeed describe a spin one we can choose our Lorentz frame so 
that 
$k^\mu = (k^+, 0, 0,\ldots , 0)$ in light-cone coordinates, 
whereupon equation (2.1.2) implies that $F^{(0)}_{+a}=0$ 
while equation (2.1.1) 
implies that $k_{[-} F^{(0)}{}_{ab]}=0$. Consequently,  the only non zero components of  the 
field strength are  $F^{(0)}{}_{- i}, \ i=1,\ldots , D-2$ subject to equation (2.1.3) and these 
we recognise as  the $D-2$ degrees of freedom of a ``spin 1". 
\par
We are now going to formulate the above conditions in an alternative manner which will 
lead to the irreducible unitary representation of $ISO(1,D-1)$ corresponding to spin one, 
but in such a way that it is manifestly $SO(1,D-1)$ covariant and also gauge invariant.  
We first observe that the conditions of equation (2.1.1) and (2.1.2) on 
$F^{(0)}{}_{a_1a_2}$ can be rewritten by defining 
$$
F^{(1)}{}_{a_1a_2\Vert  b} := \partial_b F^{(0)}{}_{a_1a_2}\;,
\eqno(2.1.4)
$$
whereupon they are equivalent to the conditions 
$$
F^{(1)}{}_{[a_1a_2\Vert  b]}=0= F^{(1)}{}_{a_1b\Vert }{}^{b}\;.
\eqno(2.1.5)$$
In the above and in what follows, 
we use conventions whereby double bars separate groups of indices 
that are subject to $GL(D)$-irreducibility conditions. 
Thus we recognise the Bianchi identity  of equations 
(2.1.1)  as just being   the requirement that  the  tensor $F^{(1)}{}_{a_1a_2\Vert  b}$ is 
$GL(D)$ irreducible. 
This is the same as stating that $F^{(1)}{}_{a_1a_2\Vert  b}$ belongs to the $GL(D)$ Young 
tableau 
$$ 
\vbox{\offinterlineskip\cleartabs
\def\hr{\vrule height .4pt width 2em}
\def\vr{\vrule height12pt depth 5pt}
\def\cc#1{\hfill#1\hfill}
\+ \hr&\hr&\cr
\+ \vr\cc{$a_1$}&\vr\cc{b}&\vr\cr
\+ \hr&\hr&\cr
\+ \vr\cc{$a_2$}&\vr&\cr
\+\hr&&\cr}\;.
\eqno(2.1.6)$$
\par
The second condition of equation (2.1.2) can be stated as that 
$F^{(1)}{}_{a_1a_2\Vert  b}$ is also a $SO(1,D-1)$ irreducible tensor.  
A Young tableau can be of $GL(D)$ or $SO(1,D-1)$ type. 
The former encodes constraints that involve the antisymmetrisation, or symmetrisation,  
of certain groups of indices, such as in the first of the equation in  (2.1.5), however, the 
latter tableau also encodes trace conditions, such as in the second equations in  (2.1.5). 
As a result, $F^{(1)}{}_{a_1a_2\vert b}$ belongs to the irreducible representation associated with 
the $SO(1,D-1)$ Young tableau given above in (2.1.6). 
The conditions encoded in the Young tableau are just those 
required to give an irreducible representation of the relevant group. 
A discussion a Young tableaux can be found in [25,26]. 
\par
We now take another  derivative and consider the quantity 
$$
F^{(2)}{}_{a_1a_2\Vert b_1b_2} := \partial_{b_2} F^{(1)}{}_{a_1a_2\Vert b_1}\;,
\eqno(2.1.7)
$$
which satisfies the conditions 
$$
F^{(2)}{}_{[a_1a_2\Vert b_1]b_2}=0= F^{(2)}{}_{a_1b\Vert }{}^b{}_{b_2}\;,
\quad F^{(2)}{}_{a_1a_2\Vert [b_1b_2]}= 0 = F^{(2)}{}_{a_1a_2\Vert b}{}^b\;.
\eqno(2.1.8)
$$
The first two conditions are obvious from the definition of equation (2.1.7) and equation 
(2.1.5) while  the last two conditions follow by substituting equation (2.1.4) into equation 
(2.1.7) and using equation (2.1.3). By considering 
$\partial^{b_1} \partial_{[ b_1} F^{(1)}{}_{b_2b_3\Vert b_4]}$, 
it follows from the constraints of equations (2.1.5) 
and  (2.1.8) that 
$$
\partial_c \partial^c F^{(1)}{}_{a_1a_2\Vert b} = 0 
= \partial_c \partial^c F^{(2)}{}_{a_1a_2\Vert b_1b_2} \;.
\eqno(2.1.9)
$$
The constraints of equation (2.1.8) are equivalent to demanding that 
$F^{(2)}{}_{a_1a_2\Vert b_1b_2}$ has the properties  associated with the 
$SO(1,D-1)$ Young tableau given by  
$$
F^{(2)}{}_{a_1a_2\Vert b_1b_2} \sim 
\vbox{\offinterlineskip\cleartabs
\def\hr{\vrule height .4pt width 2em}
\def\vr{\vrule height12pt depth 5pt}
\def\cc#1{\hfill#1\hfill}
\+ \hr&\hr&\hr&\cr
\+ \vr\cc{$a_1$}&\vr\cc{$b_1$}&\vr\cc{$b_2$}&\vr\cr
\+ \hr&\hr&\hr&\cr
\+ \vr\cc{$a_2$}&\vr&&\cr
\+\hr&&&\cr}\;.
\eqno(2.1.10)$$
\par
We now generalise the above to higher levels and define a sequence of objects 
up to level $n$:
$$
\{ F^{(p)}{}_{a_1a_2 \Vert b_1\ldots  b_{p}}\}\;, \quad p=0, 1, \ldots n\;,
\eqno(2.1.11)
$$ 
where we assume that  
$$
F^{(p)}{}_{[a_1a_2\Vert b_1]\ldots b_p}=0= F^{(p)}{}_{a_1 b\Vert }{}^b{}_{b_2\ldots b_p}\;,~
F^{(p)}{}_{a_1a_2\Vert b_1\ldots b_{p-2} c}{}^c = 0\;,  \quad p=0, 1, \ldots n\;,
\eqno(2.1.12) 
$$
and
$$
F^{(p)}{}_{a_1a_2\Vert b_1\ldots b_p}= F^{(p)}{}_{a_1a_2\Vert  (b_1\ldots b_p )}, \quad p=0,1,\ldots , n\;.
\eqno(2.1.13)$$ 
 Proceeding  to the next level $n+1$ we  define 
$$
F^{(n+1)}{}_{a_1a_2 \Vert b_1\ldots  b_{n+1}} :=  
\partial _{b_{n+1}} F^{(n)}{}_{a_1a_2\Vert b_1\ldots b_n}\;.
\eqno(2.1.14)
$$
It is now straightforward to show that $F^{(n+1)}{}_{a_1a_2 \Vert b_1\ldots  b_{n+1}} $ 
obeys equations (2.1.12) and (2.1.13) but with  $p=n+1\,$. 
Thus by induction we have an infinite set of objects which obey the constraints of 
equations (2.1.12)  and (2.1.13) for all $p\,$. 
\par
To summarise, one has a description of ``spin one'' in $D$ dimensions in terms of an infinite the set 
of objects 
$$
{\cal W} =\{ F^{(n)}{}_{a_1a_2 \Vert b_1\ldots b_n} \sim    \vbox{\offinterlineskip\cleartabs
\def\hr{\vrule height .4pt width 2em}
\def\vr{\vrule height12pt depth 5pt}
\def\cc#1{\hfill#1\hfill}
\+ \hr&\hr&\hr&\hr&\cr
\+ \vr\cc{$a_1$}&\vr\cc{$b_1$}&\vr$\ \ldots$ &\vr \cc{$b_n$}&\vr\cr
\+ \hr&\hr&\hr&\hr&\cr
\+ \vr\cc{$a_2$}&\vr&&&\cr
\+\hr&&&&\cr} , \quad n=0,1,\ldots\}
\eqno(2.1.15)$$
which are related by equation (2.1.14) and which are subject to the 
constraints that are encoded in  the $SO(1,D-1)$ Young tableau.  
\par 
The discussion   above  of all the higher level objects  may seem at first sight as a bit 
redundant,  but it has an important interpretation.  The objects of equation (2.1.15) carry
Wigner's unitary irreducible representation of $ISO(1,D-1)$  which corresponds to 
``spin one''.  Indeed, there exists a map from Wigner's unitary irreducible representation of $ISO(1,D-1)$ for ``spin one'' where all states are labelled by momentum 
and polarisation tensors, to $\cal W$, where the states are labelled by Lorentz tensors. 
The action of the Lorentz generators is as usual while  the translations acts as 
$$
P_c( F^{(n)}{}_{a_1a_2\Vert b_1\ldots b_n} )=  F^{(n+1)}{}_{a_1a_2\Vert b_1\ldots b_n c}\;.
\eqno(2.1.16)
$$
The reader may verify that ${\cal W}$  does indeed carry a representation of $ISO(1,D-1)$. We note that this representation is not irreducible, as it contains infinitely many ideals 
${\cal W}_{n_{0}}$, 
namely the modules obtained by truncating the level $n$ to any minimum value $n_{0}\,$. 
As we shall see below, it is nevertheless possible to reconstruct $\cal W$ from any ideal 
${\cal W}_{n_{0}}$ by integration with suitable boundary conditions imposed, 
conditions that we shall leave unspecified below for the sake of simplicity. 
\par
As we have mentioned, the advantage of using the above representation ${\cal W}$
and its generalisations to particles of other spin,  is that  it  is manifestly Lorentz covariant  
and in the massless case also gauge invariant and so it does not require a particular 
representation in terms of a gauge potential and its associated gauge transformations. This 
will prove key in what follows. The  representations ${\cal W}$, and its generalisations are   
equivalent to the formulation of these representations  given by the Wigner method of 
induced representations [16,17]. 
\par
The discussion above was pedagogical but to some extent a simplified account  using just 
ideas that are universally known.  In fact, the procedure is best understood from a slightly 
different and more abstract viewpoint. We should start from the beginning with the fully 
indecomposable $ISO(1,D-1)$ representation of  equation (2.1.15),  
the fields of which by definition  are subject to the $SO(1,D-1)$ conditions encoded in the 
Young tableaux, and related by the derivative condition of equation (2.1.14). 
As should be the case for this  representation, these equations imply 
the on-shell dynamics. This should be apparent from the above, for example the 
constraints on $F^{(1)}{}_{a_1a_2\Vert b}$ and the fact that 
$\partial_b F^{(1)}{}_{a_1a_2}=F^{(1)}{}_{a_1a_2\Vert b}$ implies the usual Bianchi and 
equation of motion for a spin one particle. 
The above Lorentz-covariant method of describing the particle states  is an example of what is called the {\it unfolded} description of field theory dynamics and was initiated by 
M. Vasiliev, see [19] and references therein.
This  is a formulation of the dynamics by a set of first order differential equations; in this 
case equations (2.1.14) for all $n$ together with the constraints just discussed.  
As we said, unfolded formulation plays the central role in nonlinear higher spin gravity theories.
\par
The representation ${\cal W}$ of equation (2.1.15) contains the  field $F^{(0)}{}_{a_1a_2}$ 
and all its derivatives and one can think of this as the field and all its derivatives at a given 
space-time point. Using Taylor's theorem, we then know the fields at all 
space-time points as the coefficients in the expansion are the just mentioned quantities. 
\par
It is clear from the above construction that if we have the representation ${\cal W}$  up to 
level $n$ then we can, by acting with space-time derivatives, construct all the higher level 
elements in the representation;  indeed this is what we did above. However, it is also 
possible to reconstruct $\cal W$ if we have all the elements at, and above,  
any given level $n$, which we denoted ${\cal W}_{n}$ above. 
The reconstruction is possible by using the Poincar\'e lemma. 
The fact that integration is required is to be expected, as $\cal W$ is a fully indecomposable 
representation.
Let us consider $ F^{(p)}{}_{a_1a_2| b_1\ldots  b_p}, \quad p\ge n\,$, which is subject to all the 
constraints dictated by its $SO(1,D-1)$ Young tableau of equation (2.1.15). 
These, in particular,  imply that 
$$
\partial _{[b_{n+1}|} F^{(n)}{}_{a_1a_2\Vert b_1\ldots | b_n]}=0 \;.
\eqno(2.1.17)
$$
As a result one can deduce, using the usual Poincar\'e lemma,  that there exists an object $F^{(n-1)}{}_{a_1a_2\Vert b_1\ldots  b_{n-1}}$ such that 
$$
F^{(n)}{}_{a_1a_2\Vert b_1\ldots  b_n}= \partial _{b_{n}}  F^{(n-1)}{}_{a_1a_2\Vert b_1\ldots  b_{n-1}}\;.
\eqno(2.1.18)$$
From the fact that $F^{(n)}{}_{a_1a_2\Vert b_1\ldots  b_n}$ satisfies the  algebraic 
constraints associated with  its $SO(1,D-1)$ Young tableau, it follows that 
$F^{(n-1)}{}_{a_1a_2\Vert b_1\ldots b_{n-1}}$ obeys the analogous constraints. 
Proceeding in this way we reconstruct the representation down to level zero. 
\par
It is instructive to find the degrees of freedom contained in the higher level elements of the 
representation ${\cal W}$. Let us consider $F^{(1)} _{a_1a_2\Vert b}$ which is subject to 
the constraints of equation (2.1.8)  in conjunction with equation (2.1.7), but not its 
connection to level zero, that is, to $F^{(1)}_{a_1a_2\Vert b}$ of equation (2.1.4).  
These differential constraints imply that $F^{(1)}_{a_1a_2\Vert b}$ is divergenceless and curl-free
on its two sets of indices, and as a result $F^{(1)} _{a_1a_2\Vert b}$ is harmonic. 
One goes to momentum space and takes $k^\mu = (k^+, 0, 0,\ldots , 0)\,$, 
as before, and finds that the last three  equations 
imply that the indices $a_1, a_2$ and $b$  cannot take the value $+$. 
The curl-free equations then imply that the only non-zero components are  
$F^{(1)} _{-i\Vert-}, \quad i=1,\ldots , D-2\,$.  
Hence we find it contains the required $D-2$  degrees of freedom. 
A similar analysis at level $n$ implies that the only non-zero 
components of $F^{(n)}{}_{a_1a_2\Vert b_1\ldots  b_n}$ 
are $F^{(n)} _{-i\Vert-\dots -}\,$.  We note that, in the chosen 
Lorentz frame, all the non-vanishing components are related by 
$F^{(n)} _{-i\Vert-\dots -} = k_{-}\ldots k_{-}F^{(0)}_{-i}\,$ and  
reproduce all the Taylor coefficients of an on-shell Maxwell field at any given point, 
therefore allowing to reconstruct the field in the neighbourhood of that point.   
\medskip
{\bf 2.2 Dualities and Gauge potentials} 
\medskip 
The representation ${\cal W}$ of $ISO(1,D-1)$ describes the states of a 
spin one in a way that is manifestly gauge invariant, since it is constructed from field strengths and 
their derivatives. We now consider what gauge potentials are implied by this 
representation. We begin at the lowest level. Every theorist knows that the Bianchi 
identity of equation (2.1.1) can be solved in terms of a gauge potential $A^{(0)}_a$ as 
$$
F^{(0)}{}_{a_1a_2}= 2\,\partial _{[a_1} A^{(0)}{}_{a_2]}\;,
 \eqno(2.2.1)
 $$
with the usual gauge symmetry $\delta A_a= \partial_a \Lambda$. 
\par
However, we are free to choose which of the equations in the representation we would like 
to solve and we can equally well choose to solve equation (2.1.2) even though this is 
usually thought of as the equation of motion. To achieve this we define 
$$
G_{a_1\ldots a_{D-2}}={1\over 2} \, 
\epsilon _{ a_1 \ldots a_{D-2}}{}^{ b_1 b_{2}}F_{b_1b_2}\;,
\eqno(2.2.2)
$$ 
whereupon the equation (2.1.2) becomes 
$$
 \epsilon ^{a_1a_2b_1\ldots b_{D-2}  } \partial _{a_2} G_{b_1\ldots b_{D-2}}=0 \;,
 \eqno(2.2.3)
 $$
with the solution 
$$
 G_{b_1\ldots b_{D-2}}= \partial_{[ b_1} A^{(0)}{}_{b_2 \ldots b_{D-2}]}\;,
\eqno(2.2.4)
$$
that is, in terms of a gauge field $ A^{(0)}_{b_1 \ldots b_{D-3}}$ with the gauge symmetry 
$$ \delta_{\lambda} A^{(0)}_{b_1 \ldots b_{D-3}} = \partial_{[ b_1}  
\lambda_{b_2 \ldots b_{D-3}]}\;.
$$ 
\par
However as we can reconstruct $\cal W$ from ${\cal W}_{n}$ by integration, 
we choose to carry out a duality at level $n$ rather than those at level zero. 
Let us first consider level one and define 
$$
G^{(1)}{}_{ c_1\ldots c_{D-1}\Vert a_1a_2} :=  \epsilon _{ c_1\ldots c_{D-1}b} 
F^{(1)}{}_{a_1a_2\Vert}{}^{b}\;.
\eqno(2.2.5)
$$ 
It is straightforward to verify, using equation (2.1.2),  that is, the trace constraint 
$F^{(1)}{}_{ab\Vert }{}^{b}$ $= \partial^b F_{ab}=0$,   that 
$$
G^{(1)}{}_{ [c_1\ldots c_{D-1}\Vert a_1]a_2}=0 \;,
\eqno(2.2.6)
$$
Thus, $G^{(1)}{}_{ c_1\ldots c_{D-1}\Vert a_1a_2}$ sits inside the irreducible representation 
of $GL(D)$ that transforms according to the following  $GL(D)$ Young tableau
$$
G^{(1)}{}_{ c_1\ldots c_{D-1}\Vert a_1a_2}\sim 
\vbox{\offinterlineskip\cleartabs
\def\hr{\vrule height .4pt width 2.4em}
\def\vr{\vrule height15pt depth 5pt}
\def\cc#1{\hfill#1\hfill}
\+ \hr&\hr&\cr
\+ \vr\cc{$c_1$}&\vr\cc{$a_1$}&\vr\cr
\+ \hr&\hr&\cr
\+ \vr\cc{$c_2$}&\vr\cc{$a_2$}&\vr\cr
\+\hr&\hr&\cr
\+ \vr\cc{$\vdots$}&\vr&\cr
\+ \hr&&\cr
\+ \vr\cc{$c_{D-1}$}&\vr&\cr
\+ \hr& &\cr}\;.
\eqno(2.2.7)$$
However, it is easy to see that $G^{(1)}{}_{ c_1\ldots c_{D-2}b\Vert }{}^{b}{}_{a_2}\not= 0$ 
and so the constraints on this object are not those of an $SO(1,D-1)$ Young tableau, 
which are single traceless by definition. 
It does however satisfy a higher order trace condition, and using 
 the first equation of (2.1.5),  one finds that 
$$
G^{(1)}{}_{ a_1a_2 c_1\ldots c_{D-3}\Vert }{}^{a_1a_2}=0\;.
\eqno(2.2.8)
$$
We can think of equation (2.2.8) as the equation of motion and equation (2.2.6) as the 
Bianchi identity for the particle when written in terms of 
$G^{(1)}{}_{ c_1\ldots c_{D-1}\Vert a_1a_2}$. Indeed we will show below that equation (2.2.8)    follows 
from extremising an action. 
We note the usual  interchange of equation of motion and Bianchi identity under a duality 
transformation. 
\par
We next consider what derivative constraints $G^{(1)}{}_{ c_1\ldots c_{D-1}\Vert a_1a_2}$ satisfies. Using equation (2.1.8) and the definition (2.2.5) we find that 
$$
\partial _{[a_1} G^{(1)}{}_{ c_1\ldots c_{D-1}\Vert a_2a_3]}= 0 =
\partial _{[c_1} G^{(1)}{}_{ c_2\ldots c_{D}]\Vert a_1a_2}\;,
\eqno(2.2.9)$$
and 
$$
\partial ^{e} G^{(1)}{}_{ c_1\ldots c_{D-1}\Vert ea}=0=
\partial ^{d} G^{(1)}{}_{ d c_2\ldots c_{D-1}\Vert a_1a_2}\;.
\eqno(2.2.10)
$$
Equations (2.2.9) can be summarised by defining  
$$
G^{(2)}_{ c_1\ldots c_{D-1}\Vert a_1a_2\Vert b}:= \partial _{b} G^{(1)}{}_{ c_1\ldots c_{D-1}\Vert a_2a_3}\;,
\eqno(2.2.11)
$$
and demanding that  it belong to the $GL(D)$ Young tableau 
$$
\vbox{\offinterlineskip\cleartabs
\def\hr{\vrule height .4pt width 2.5em}
\def\vr{\vrule height15pt depth 5pt}
\def\cc#1{\hfill#1\hfill}
\+ \hr&\hr&\hr&\cr
\+ \vr\cc{$c_1$}&\vr\cc{$a_1$}&\vr\cc{b}&\vr \cr
\+ \hr&\hr&\hr&\cr
\+ \vr\cc{$c_2$}&\vr\cc{$a_2$}&\vr&\cr
\+\hr&\hr&&\cr
\+ \vr\cc{$\vdots$}&\vr&&\cr
\+ \hr&&&\cr
\+ \vr\cc{$c_{D-1}$}&\vr&&\cr
\+ \hr& &&\cr}\;.
\eqno(2.2.12)
$$
The tensor $G^{(2)}_{ c_1\ldots c_{D-1}\Vert a_1a_2\Vert b}$ 
satisfies trace conditions inherited from equation (2.2.10) and an obvious second order 
trace condition inherited from (2.2.8). Because of the latter condition, 
these are not the single-trace conditions associated with an $SO(1,D-1)$-irreducible Young tableau. 
\par
Before introducing potentials, 
we can proceed a but further as we did for $F_{a_1a_2}$ to construct an 
infinite dimensional 
representation based on  $G^{(1)}_{ c_1\ldots c_{D-1}\Vert a_1a_2}$ by 
defining 
$$
G^{(n+1)}_{ c_1\ldots c_{D-1}\Vert a_1a_2\Vert b_1\ldots b_{n+1}}:= \partial _{b_{n+1}} G^{(n)}{}_{ c_1\ldots c_{D-1}\Vert a_2a_3 \Vert b_1\ldots b_{n}}\;,
\eqno(2.2.13)
$$
and find the constraints that the new objects  satisfies. 
However, we will not pursue this further in this work.  
\par
We now choose to regard   equations (2.2.9) as Bianchi identities and so solve these 
instead of the Bianchi identity  at level zero. Using the generalised Poincar\'e lemma 
spelled out in section 5 of [28], we find that 
$$
G^{(1)}{}_{ c_1\ldots c_{D-1}\Vert}{}^{ a_2a_3}= \partial ^{[a_1}\partial _ {[c_1}{}
A^{(1)}{}_{ c_2\ldots c_{D-1}]\Vert }{}^{a_2]} \;.
\eqno(2.2.14)
$$
Thus we find a description in terms of a gauge field 
$A_{ a_1\ldots a_{D-2}\Vert b}$ which satisfies the $GL(D)$ irreducibility condition 
$$
A^{(1)}_{ [a_1\ldots a_{D-2}\Vert b]}=0\;.
\eqno(2.2.15)
$$
The expression for the field strength in terms of the gauge field can be written in the form 
of a Young tableau as follows 
$$
\vbox{\offinterlineskip\cleartabs
\def\hr{\vrule height .4pt width 2.5em}
\def\vr{\vrule height15pt depth 5pt}
\def\cc#1{\hfill#1\hfill}
\+ \hr&\hr&\cr
\+ \vr\cc{$\partial_{c_1}$}&\vr\cc{$\partial_{a_1}$}&\vr \cr
\+ \hr&\hr&\cr
\+ \vr\cc{$c_2$}&\vr\cc{${a_2}$}&\vr\cr
\+\hr&\hr&\cr
\+ \vr\cc{$\vdots$}&\vr&\cr
\+ \hr& &\cr
\+ \vr\cc{${c_{D-1}}$}&\vr &\cr
\+ \hr& &\cr}\;.
\eqno(2.2.16)
$$
\par
 The fields strength $G^{(1)}{}_{ c_1\ldots c_{D-1}\Vert  a_2a_3}$ is invariant 
 under the following gauge transformations featuring two independent 
$GL(D)$-irreducible gauge parameters $\lambda^{(1)}{}_{a_1a_2\ldots a_ {D-3} \Vert b}$ 
and $\lambda^{(2)}{}_{a_1a_2\ldots a_ {D-2}}\,$: 
$$
\delta_{\lambda} A^{(1)}{}_{ a_1\ldots a_{D-2}\Vert b}
 = (D-2)\,\partial_{ [a_1}\lambda^{(1)}{}_{a_2\ldots a_{D-2}]\Vert b } 
 + \partial_{b}\lambda^{(2)}{}_{a_1\ldots a_ {D-2} } + (-1)^{D-1} 
\partial_{[a_{1}}\lambda^{(2)}{}_{a_2\ldots a_ {D-2}] b}\;.
\eqno(2.2.17)
$$
These two transformations can be represented by the tableaux  
$$
\vbox{\offinterlineskip\cleartabs
\def\hr{\vrule height .4pt width 2.6em}
\def\vr{\vrule height15pt depth 5pt}
\def\cc#1{\hfill#1\hfill}
\+ \hr&\hr&\cr
\+ \vr\cc{$a_1$}&\vr\cc{$ b$}&\vr \cr
\+ \hr&\hr&\cr
\+ \vr\cc{$a_2$}&\vr&\cr
\+ \hr&&\cr
\+ \vr\cc{$\vdots$}&\vr&\cr
\+ \hr& &\cr
\+ \vr\cc{$\partial_{a_{D-2}}$}&\vr &\cr
\+ \hr& &\cr}\qquad \qquad , \qquad \qquad
\vbox{\offinterlineskip\cleartabs
\def\hr{\vrule height .4pt width 2.5em}
\def\vr{\vrule height15pt depth 5pt}
\def\cc#1{\hfill#1\hfill}
\+ \hr&\hr&\cr
\+ \vr\cc{${a_1}$}&\vr\cc{$\partial_{b}$}&\vr \cr
\+ \hr&\hr&\cr
\+ \vr\cc{$a_2$}&\vr&\cr
\+\hr&&\cr
\+ \vr\cc{$\vdots$}&\vr&\cr
\+ \hr& &\cr
\+ \vr\cc{${a_{D-2}}$}&\vr &\cr
\+ \hr& &\cr}\;.
\eqno(2.2.18)$$
\par
We now turn to the case of dualising  the  object in the irreducible 
representation of $SO(1,D-1)$ given in equation (2.1.15)  at level $n\,$. 
To this end, we need a more streamlined index notation. 
We denote $A_{a[n]}\equiv A_{[a_1\ldots a_n]} \equiv A_{a_1\ldots a_n}$ 
and similarly for all blocks of  antisymmetric  indices. 
Similarly, for groups of symmetric indices, we use 
$S_{a(n)}\equiv S_{(a_1\ldots a_n)} \equiv S_{a_1\ldots a_n}\,$, with
strength-one (anti)symmetrisation convention.  
Using this notation we define 
$$
G^{(n)}{}_{ c^{1}[D-1]\Vert  \ldots \Vert  c^{n}[D-1] \Vert a_1a_2} :=  
\epsilon _{c^{1}[D-1]e_{1}} \ldots 
\epsilon _{c^{n}[D-1]e_{n}} \;
F^{(n)}{}_{a_1a_2\Vert }{}^{e(n)}\;.
\eqno(2.2.19)
$$
Using equations (2.1.12) and (2.1.13)  one can show that 
$G^{(n)}{}_{ c^{1}[D-1]\Vert  \ldots \Vert  c^{n}[D-1] \Vert a_1a_2}$ 
 obeys the following over-antisymmetrisation constraints 
$$
G^{(n)}{}_{ c^{1}[D-1] | \Vert  \ldots \Vert c^{1}_D c^{i}[D-2] \Vert \ldots \Vert c^{n}[D-1] \Vert a_1a_2}=0
= G^{(n)}{}_{ c^{1}[D-1]\Vert  \ldots \Vert c^{j}[D-1] \Vert  \ldots \Vert c^{n}[D-1] \Vert c^{j}_D a_2}\;,
$$
$$
i \in \{ 2,\ldots, n\}\;, \quad j \in \{ 1,2,\ldots, n\}\;.
\eqno(2.2.20)
$$
As a result, $G^{(n)}$ belongs to the $GL(D)$ Young tableau 
$$
\vbox{\offinterlineskip\cleartabs
\def\hr{\vrule height .4pt width 2.5em}
\def\vr{\vrule height15pt depth 5pt}
\def\cc#1{\hfill#1\hfill}
\+ \hr&\hr&\hr&\hr&\cr
\+ \vr\cc{$c^{1}_1$}&\vr\cc{$\ldots $}&\vr\cc{$c^{n}_1$}&\vr\cc{$a_1$}&\vr \cr
\+ \hr&\hr&\hr&\hr&\cr
\+ \vr\cc{$c^{1}_2$}&\vr\cc{$\ldots$} &\vr \cc{$c^{n}_2$}&\vr\cc{$a_2$}&\vr\cr
\+\hr&\hr&\hr&\hr\cr
\+ \vr\cc{$\vdots$}&\vr\cc{$\ldots$} &\vr\cc{$\vdots$}&\vr&\cr
\+ \hr&\hr&\hr&&&\cr
\+ \vr\cc{$c^{1}_{D-1}$}&\vr\cc{$\ldots$} &\vr\cc{$c^{n}_{D-1}$}&\vr&\cr
\+ \hr&\hr&\hr& &\cr}.
\eqno(2.2.21)$$
\par
It is straightforward to show that although
 $G^{(n)}{}_{ c[D-1]\Vert  \ldots \Vert  d[D-1] \Vert a_1a_2}$ does not satisfy any single 
 trace conditions it does satisfy a double and a $(D-1)\,$-trace condition which are given by 
$$
G^{(n)}{}_{ c[D-1]\Vert  \ldots \Vert  d[D-3]a_1a_2 \Vert }{}^{a_1a_2}=0=
G^{(n)}{}_{ c[D-1]\Vert  \ldots \Vert }{}^{ c[D-1]}{}_{ \Vert a_1a_2}\;.
\eqno(2.2.22)
$$
\par
The dynamics of the ``spin one'' when expressed in terms of the field strength 
$\quad$ $G^{(n)}{}_{ c[D-1]\Vert  \ldots \Vert  d[D-1] \Vert a_1a_2}$ is given by equations 
(2.2.20) and (2.2.22) which  replace equations (2.1.1) and (2.1.2) of the usual formulation 
in terms of the field strength $F^{(0)}_{a_1a_2}$. We can think of equation (2.2.20) as generalised 
Bianchi identities at level $n-1$ and equations (2.2.22), which involve traces,  as equations of motion.
\par
Using the equations (2.2.19) and  equation (2.1.13),  we find that the field 
strength $G^{(n)}{}_{ c[D-1]\Vert  \ldots \Vert  d[D-1] \Vert a_1a_2}$  also obeys the 
curl-free conditions 
$$
\partial_{c} G^{(n)}{}_{ c[D-1] \Vert  \ldots \Vert  d[D-1] \Vert a_1a_2}=0
= \partial_{a_1}G^{(n)}{}_{ c[D-1]\Vert  \ldots \Vert  d[D-1] \Vert a_2a_3}\;,
\eqno(2.2.23)
$$
as well as the divergence-free conditions
$$
\partial^{e} G^{(n)}{}_{ e c[D-2]\Vert  \ldots \Vert  d[D-1] \Vert a_1a_2}=0
= \partial^{e }G^{(n)}{}_{ c[D-1]\Vert  \ldots \Vert  d[D-1] \Vert e a}\;.
\eqno(2.2.24)
$$
We note that (2.2.23)  will be the generalised Bianchi identities at the level $n$
while the equation of motion will not be (2.2.24) but instead the higher-trace constraints (2.2.20).
\par
We  find the gauge potential at level $n$ by  applying  the generalised Poincar\'e lemma 
[28] to equation (2.2.23). The result is 
$$
G^{(n)}{}_{ c^{1}[D-1]]\Vert  \ldots \Vert  c^{n}[D-1] \Vert}{}^{a_1a_2}\; = \; \partial^{[a_1 }
\partial_{[c^{1}_1} \ldots \partial _{[c^{n}_1} A^{(n)}{}_{c^{1}[D-2]]\Vert  \ldots \Vert  c^{n}[D-2]] \Vert }{}^{a_2]} 
\;,
\eqno(2.2.25)
$$
where the 
gauge potential $A^{(n)}{}_{c[D-2]\Vert  \ldots \Vert  d[D-2] \Vert a}$ is an irreducible 
tensor of $GL(D)$ and so obeys the constraints 
$$
A^{(n)}{}_{c[D-2]\Vert  \ldots \Vert  [f[D-2] \Vert a]} = 0 
= A^{(n)}{}_{[c[D-2]|\Vert  \ldots \Vert | f_{1}] | f[D-3] \Vert a}\;.
\eqno(2.2.27)
$$
As a result $A^{(n)}_{c[D-2]\Vert  \ldots \Vert  f[D-2] \Vert a}$ belongs to the 
$GL(D)$ Young tableau
$$
\vbox{\offinterlineskip\cleartabs
\def\hr{\vrule height .4pt width 2.5em}
\def\vr{\vrule height15pt depth 5pt}
\def\cc#1{\hfill#1\hfill}
\+ \hr&\hr&\hr&\hr&\cr
\+ \vr\cc{$c_1$}&\vr\cc{$\ldots $}&\vr\cc{$f_1$}&\vr\cc{$a$}&\vr \cr
\+ \hr&\hr&\hr&\hr&\cr
\+ \vr\cc{$c_2$}&\vr\cc{$\ldots$} &\vr \cc{$f_2$}&\vr&\cr
\+\hr&\hr&\hr&&\cr
\+ \vr\cc{$\vdots$}&\vr\cc{$\ldots$} &\vr\cc{$\vdots$}&\vr&\cr
\+ \hr&\hr&\hr&&\cr
\+ \vr\cc{$c_{D-2}$}&\vr\cc{$\ldots$} &\vr\cc{$f_{D-2}$}&\vr&\cr
\+ \hr&\hr&\hr& &\cr}\;,
\eqno(2.2.28)$$  
The potential has no trace constraint and has gauge symmetries involving 
two gauge parameters, 
$\Lambda^{(n,1)}_{[D-2,\ldots,D-2]}$ and $\Lambda^{(n,2)}_{[D-2,\ldots,D-2,D-3,1]}$. 
\par
We note that the field strength 
$G^{(n)}{}_{ c[D-1]]\Vert  \ldots \Vert  d[D-1] \Vert a_1a_2}$ 
involves $n+1$ space-time derivatives, instead of the more familiar two derivatives.  
The expression of the field strength 
$G^{(n)}{}_{ c[D-1]]\Vert  \ldots \Vert  f[D-1] \Vert a_1a_2}$ in terms of the gauge field 
$A^{(n)}{}_{c[D-2]\Vert  \ldots \Vert  f[D-2] \Vert a}$ given in equation (2.2.25) can be expressed 
in Young tableau language as 
$$
\vbox{\offinterlineskip\cleartabs
\def\hr{\vrule height .4pt width 2.5em}
\def\vr{\vrule height15pt depth 5pt}
\def\cc#1{\hfill#1\hfill}
\+ \hr&\hr&\hr&\hr&\cr
\+ \vr\cc{$\partial_{c_1}$}&\vr\cc{$\ldots $}&\vr\cc{$\partial_{f_1}$}&\vr\cc{$\partial_{a_1}$}&\vr \cr
\+ \hr&\hr&\hr&\hr&\cr
\+ \vr\cc{$c_2$}&\vr\cc{$\ldots$} &\vr \cc{$f_2$}&\vr\cc{${a_2}$}&\vr\cr
\+\hr&\hr&\hr&\hr&\cr
\+ \vr\cc{$\vdots$}&\vr\cc{$\ldots$} &\vr\cc{$\vdots$}&\vr&\cr
\+ \hr&\hr&\hr&&&\cr
\+ \vr\cc{$c_{D-2}$}&\vr\cc{$\ldots$} &\vr\cc{$f_{D-2}$}&\vr&\cr
\+ \hr&\hr&\hr& &\cr
\+ \vr\cc{${c_{D-1}}$}&\vr\cc{$\ldots $}&\vr\cc{${f_{D-1}}$} &\vr\cr    
\+ \hr&\hr&\hr& & \cr}\;.
\eqno(2.2.28)$$
\par
The fields strength $G^{(n)}{}_{ c[D-1]]\Vert  \ldots \Vert  d[D-1] \Vert a_1a_2}$ when 
expressed in terms of its gauge potential in equation  (2.2.25) automatically obeys the 
Bianchi identities of equations (2.2.20) and (2.2.23). However, it also obeys equation (2.2.22)  
which is the equation of motion for the gauge field, and we stress that it contains  $n+1$ space-time 
derivatives.
The presence of higher space-time derivatives is characteristic of the equations of motion  
of higher spin fields and indeed  of  any mixed symmetry fields, when formulated in terms of curvatures.  
For a recent discussion and references along those lines, see [27].  
\par
Thus we have found  that there is an infinite number of ways of representing 
the particle states of  ``spin one" corresponding to the existence of an infinite  number of 
different possible gauge potentials arising from the infinite number of ways of 
dualising the field strength  and its descendants that occur in the Lorentz-covariant unfolded 
module $\cal W$. 
We repeat that the latter module carries the irreducible unitary 
representation of $ISO(1,D-1)$ that describes the states of the ``spin one". 
\par
 We now turn to a key point of this paper, which is the duality relations between the  first-order derivatives of the potentials. To obtain these, we will first obtain  
duality relation between the $(n+1)$-derivative field strengths 
when they are expressed in terms of their respective gauge potentials. 
We begin at  level zero and in particular 
equation (2.2.2) which now relates the gauge field $A_a$ to the gauge field $A_{a[D-3]}$. 
This duality relation is of a familiar type in so much as it relates equations of motion to 
Bianchi identities. However, once we have substituted in the gauge potentials the Bianchi identities hold automatically and so the relations imply the equations of motion.   
\par
We now consider  duality relation at level $n=1$,  which was  given in equation (2.2.5) and   
that can be written as  
$$
\partial_{[a_1}\partial_{[c_1}A^{(1)}{}_{c[D-2]]\Vert}{}_{a_2]} =  \epsilon _{c[D-1]b}  
\partial^b\partial_{[a_1}A^{(0)}{}_{a_2]} \;.
\eqno(2.2.29)$$
We note that the Bianchi identities of equations of  (2.1.1), or equivalently the first equation in (2.1.5), and 
equation (2.2.6)  are automatically  satisfied. However as the duality relation  interchange Bianchi 
identities with equations of motion for the two fields we find that equation (2.2.29) automatically imposes 
the equations of motion of the two fields, namely, the second equation of (2.1.5) and equation (2.2.8), 
{\it i.e.}
$$
\partial ^b F_{ab}=0
\eqno(2.2.30)
$$
and 
$$
\partial^{a_1}\partial_{[a_1}A^{(1)}{}_{a_2 c[D-3] ]\Vert }{}^{a_2}=0\;.
\eqno(2.2.31)
$$
Another, more direct way of getting these two equations directly from (2.2.29)
is to antisymmetrise all the $c$ indices together with $a_{1}$ of that equation, which gives 
(2.2.30), or to take its double trace, which gives of (2.2.31). 
\par
We now consider the  duality relations at higher levels. 
We begin with the relation (2.2.19) but write it in the form 
$$
G^{(n)}{}_{ c[D-1]\Vert  \ldots \Vert  b[D-1] \Vert a_1a_2}=
\epsilon _{ b[D-1]}{}^{f} G^{(n-1)}{}_{ c[D-1]\Vert  \ldots  \Vert a_1a_2\Vert f}=
$$
$$
=\epsilon _{ b[D-1]}{}^{f} \partial_f G^{(n-1)}{}_{ c[D-1]\Vert  \ldots  \Vert a_1a_2}\;,
\eqno(2.2.32)
$$
which relates field strengths at adjacent levels. We now examine the effect of imposing the 
Bianchi identities and equations of motion on each of these field strengths without 
assuming that they are given in terms of   the gauge fields. For indices that are not 
involved in the duality, that is do not occur on the epsilon symbol, the constraints on one 
side of the equation obviously hold on the other side. As a result we now consider the 
constraints   that involve  indices that occur  in the duality. The  Bianchi identities of 
$G^{(n)}{}_{ c[D-1]\Vert  \ldots \Vert  b[D-1] \Vert a_1a_2}$  of equation (2.2.20) imply 
the trace conditions for $G^{(n-1)}{}_{ c[D-1]\Vert  \ldots  \Vert a_1a_2\Vert f}$, namely 
$$
G^{(n)}{}_{ c[D-1]\Vert  \ldots \Vert [ b[D-1] \Vert a_1]a_2}=0\quad \Longleftrightarrow \quad 
G^{(n-1)}{}_{ c[D-1]\Vert  \ldots  \Vert a_1b\Vert }{}^{b}=0\;,
\eqno(2.2.33)
$$
and 
$$
G^{(n)}{}_{ [c[D-1]\Vert  \ldots \Vert  b_1]b[D-2] \Vert a_1a_2}=0
\quad \Longleftrightarrow \quad 
G^{(n-1)}{}_{e c[D-2]\Vert  \ldots  \Vert a_1a_2\Vert }{}^{e}=0\;.
\eqno(2.2.34)
$$
Conversely the Bianchi identities of $G^{(n-1)}{}_{ c[D-1]\Vert  \ldots  \Vert a_1a_2\Vert f}$ 
imply the trace conditions of $G^{(n)}{}_{ c[D-1]\Vert  \ldots \Vert  b[D-1] \Vert a_1a_2}$ of 
equation (2.2.22), namely 
$$
G^{(n-1)}{}_{ [c[D-1]|\Vert  \ldots  \Vert a_1a_2\Vert |f]}=0\quad \Longleftrightarrow \quad   G^{(n)}{}_{ c[D-1]\Vert  \ldots \Vert  }{}^{c[D-1]}{}_{ \Vert }{}_{a_1a_2}=0\;,
\eqno(2.2.35)
$$
and 
$$
G^{(n-1)}{}_{ c[D-1]\Vert  \ldots  \Vert [a_1a_2\Vert f]}=0\quad \Longleftrightarrow \quad 
G^{(n)}{}_{ c[D-1]\Vert  \ldots \Vert  a_1a_2 b[D-3] \Vert}{}^{ a_1a_2}=0\;.
\eqno(2.2.36)
$$
\par
Substituting for the gauge field in the duality relation of equation  (2.2.32) yields 
$$
\partial^{[a_1}\partial_{[c_1} \ldots \partial_{[b_1}
A^{(n)}{}_{b[D-2]]\Vert \ldots \Vert c[D-2]]\Vert }{}^{a_2]}
= \epsilon _{ b[D-1]}{}^{f} \partial_f 
\partial^{[a_1}\partial_{[c_1} \ldots 
A^{(n-1)}{}_{\ldots \Vert c[D-2]]\Vert}{}^{a_2]}\;.
\eqno(2.2.37)
$$
Once we have substituted the gauge fields in the field strengths,  the Bianchi identities, which occur on the left hand-sides of equations (2.2.33-36),  are  
automatically satisfied and as a result the trace conditions on the dual field strengths  are now enforced. In particular,  examining equation (2.2.35) and (2.2.36),  we now find 
that their left-hand sides vanish automatically and so the gauge field $A^{(n-1)}_{[D-1]\Vert \ldots \Vert a}$ does not appear in this relation. Consequently,   the right-hand side of these relations  are enforced and we find that the field strength 
$G^{(n)}{}_{ c[D-1]\Vert  \ldots \Vert b[D-1] \Vert a_1a_2}$ satisfy 
the trace conditions, which 
are  the equations of motion for the gauge field 
$A^{(n)}_{c[D-2]\Vert \ldots \Vert b[D-1]\Vert a}$. 
\par
We note that the field strength is symmetric under the exchanges of its columns 
of $D-1$ indices and so the trace 
conditions hold on all these columns and not just for the the ones displayed above. Hence the 
duality condition of equation (2.2.37) implies the equation of motion for the ``spin one" in 
the formulation with the level $n$  gauge field. Examining equations (2.2.33) and (2.2.34) 
we find a similar  conclusion but now the gauge field 
$A^{(n)}_{c[D-2]\Vert \ldots b[D-1]\Vert a}$ is eliminated and we have the equation of 
motion for the gauge field $A^{(n-1)}_{c[D-2]\Vert \ldots b[D-2]\Vert a}$ field. 
\par
Equations  (2.2.37) can be thought of as an infinite set of duality relations for 
$n=1,2,\ldots $, the first of which is given in equation (2.2.29). We note that  they involve 
ever increasing  numbers of space-time derivatives as $n$ increases. However, as we now 
show we can integrate these equations such that they only involve a single space-time 
derivative. At the lowest level we find,  integrating equation (2.2.29), that 
$$
\partial_{c}A^{(1)}{}_{c[D-2]\Vert }{}_{a} = \epsilon _{c[D-1]b}\; \partial^{b}A^{(0)}{}_{a} 
+ \partial_{a}\Xi_{c[D-1]}\;,
\eqno(2.2.38)$$
where the last term is the general solution of the homogeneous equation. 
 We can rewrite equation (2.2.38)  as 
$$
\partial_{c_1}A^{(1)}{}_{c[D-2]\Vert }{}_{a} = \epsilon _{c[D-1]b}\{ \partial^{b}A^{(0)}{}_{a} 
+ \partial_{a}\Xi ^b\},
\eqno(2.2.39)$$
where $\Xi_{c[D-1]}= \epsilon _{c[D-1]b}\Xi^b$. However, since $\Xi^b$ is arbitrary, we can  
shift it as $\Xi^b\to \Xi^b -A^b$
whereupon our original equation becomes 
$$
\partial_{c}A^{(1)}{}_{c[D-2]\Vert }{}_{a} = \epsilon _{c[D-1]}{}^{b}\; \partial_{[b}A^{(0)}{}_{a]} 
+ \partial_{a}\Xi_{c[D-1]}\;,
\eqno(2.2.40)$$
We observe that the equation is now  invariant under the gauge transformations of the original gauge field $A^{(0)}{}_{a}$. 
Antisymmetrising on $\{ c_{1},\ldots,c_{D-1},a \} \,$, we find that 
$\partial_c\Xi_{c[D-1]}=0$ and so $\Xi_{c[D-1]}= \partial_c\Xi_{c[D-2]}$. 
Substituting this back in equation (2.2.38) it becomes 
$$
\partial_{c_1}A^{(1)}{}_{c[D-2]\Vert }{}_{a} = \epsilon _{c[D-1]}{}^{b}\; 2\partial_{[b}A^{(0)}{}_{a]} 
+ \partial_{a}\partial_c\Xi_{c[D-2]}\;.
\eqno(2.2.41)$$
We recognise that the presence of the last term ensures the invariance of (2.2.41) 
under the gauge transformation 
of the second type in equation (2.2.18), which acts as a shift symmetry on $\Xi_{c[D-2]}\,$.
The price for the integration is that  the equation is now gauge-invariant only at the price of 
an extra field with a shift symmetry. 
To eliminate the extra field requires 
that we differentiate and antisymmetrise with the $a$ index, so recovering the original 
relation of equation (2.2.29). 
\par
Integrating at higher levels, in particular equation (2.2.37),  we find that 
$$
\partial_{[b_1} A^{(n)}{}_{b[D-2]]\Vert c^{1}[D-2]\Vert\ldots \Vert c^{n-1}[D-2]\Vert a } 
= \epsilon _{b[D-1] f}\;\partial^f A^{(n-1)}{}_{c^{1}[D-2]\Vert \ldots \Vert c^{n-1}[D-2]\Vert a}\; + 
$$
$$
+ Y \left( \partial_{a}\Sigma_{b[D-1]|c^{1}[D-2]\Vert \ldots\Vert c^{n-1}[D-2]} + 
\partial_{c^{n-1}}\Xi_{b[D-1]|c^{1}[D-2]\Vert\ldots\Vert c^{n-1}[D-3]\Vert a} \right)\;,
$$
$$
\quad n=0,1,2,\ldots \;,\eqno(2.2.42)
$$
where $Y(\cdot)$ denotes the projection on the $GL(D)$ Young tableau 
with index structure $\{ c^{1}[D-1]\Vert \ldots \Vert c^{n-1}[D-1]\Vert a\,\}$. 
Using arguments similar to those given below equation (2.2.38) one can bring the duality relation to 
a form that is invariant under certain of the gauge transformations of the field 
$A^{(n-1)}{}_{c_{1}[D-2]\Vert \ldots \Vert c_{n-1}[D-2]\Vert a}$ and it  then holds modulo the 
remaining gauge transformations of the two fields, 
\par
Rather than constructing the infinite set of duality relations beginning  with the gauge field 
$A^{(0)}_{a}$ 
we can alternatively use the level zero  gauge field $A^{(0)}_{b_1 \ldots b_{D-3}}$ and repeat all the 
above steps. Including this step  we find a formulation of the ``spin one" field in terms of the 
following gauge fields 
$$
A_{[1]},\; A_{[D-3]}, \;A_{[D-2, 1]}, \;A_{[D-2, D-3]}, \;A_{[D-2, D-2, 1]}, 
$$
$$
A_{[D-2, D-2, D-3]},\;
\dots , \;A_{[D-2, \ldots , D-2, 1]},\; A_{[D-2,\ldots ,  D-2, D-3]},\;\ldots 
\eqno(2.2.43)$$
where the numbers shown as subscripts  between square brackets indicate the number 
of indices in each block, that is the length of columns in the corresponding Young tableau.  
\par
\medskip

Thus, in summary we have shown that  the spin one can be described by an infinite set of duality 
equations which are first order in space-time derivatives but only hold modulo certain gauge 
transformations,  One might suspect that these duality relations are invariant under an infinite duality 
symmetry, modulo the gauge transformations. Indeed one might suppose that this can be formulated as a 
non-linear realisation of an algebra with generators that carry the same indices that are those carried by 
the gauge fields but raised. 
\medskip 
\medskip
{\bf 2.3 Action principle for the dual potential $A^{(1)}_{c[D-2]\Vert a}$}
\medskip

In this section, we follow the lines sketched in [14] and give the action 
describing the 
dynamics of a Maxwell field in terms of the dual potential 
$A^{(1)}_{c[D-2]\Vert a}$ introduced
above and sometimes denoted $A_{[D-2,1]}$, for the sake of brevity. 
The way we recover the dynamics (2.2.8) is interestingly subtle. 
As explained in the context of the Fierz--Pauli theory in [14], 
the various dual actions involve more and more off-shell fields and are therefore 
less and less economical. 
The special case of spin-1 is simpler but allows us to see in a very explicit way the 
mechanism whereby the extra off-shell fields disappear from the dynamics on shell.  
\medskip 

We start, as it should, with the Maxwell action, and integrate by part: 
$$
S[A] = -{1 \over 2} \int d^{D}x\, \partial_{a}A_{b}(\partial^{a}A^{b} - \partial^{b}A^{a} )
= -{1 \over 2} \int d^{D}x\, (\partial_{a}A_{b} \partial^{a}A^{b} -\partial_{a}A^{a}\partial_{b}A^{b})\;,
\eqno(2.3.1)
$$
dropping the boundary term. 
Introducing the following parent action 
$$
S[Y,P] = \int d^{D}x\, (P_{a\vert}{}^{b} \partial_{c}Y^{ca\vert}{}_{b}  - {1 \over 2} 
P^{a}{}_{|b}P^{b}{}_{|a} + {1 \over 2} P^{a}{}_{|a}P^{b}{}_{|b} )\;
\eqno(2.3.2)
$$
that features two fields, $Y^{ca\vert}{}_{b} = - Y^{ac\vert}{}_{b}$ and $P_{a\vert}{}^{b}\,$, 
we reproduce the original action (2.3.1) upon extremising $S[Y,P]$ with respect to $Y$:
$$
\partial_{[c}P_{a]\vert}{}^{b} = 0 \quad \Leftrightarrow \quad P_{a\vert}{}^{b} = \partial_{a}A^{b}\;,
\eqno(2.3.3)
$$
and plugging back into (2.3.2). On the other hand, $P_{a\vert}{}^{b} $ is an auxiliary field, 
so 
that extremising the action with respect to it enables one to express it 
in terms of the $Y$ field: 
$$
P_{b\vert}{}^{a} = \partial_{c}Y^{ca\vert}{}_{b} - {1\over (D-1)}\, \delta^{a}_{b}\,
\partial_{c}Y^{cd\vert}{}_{d}\;.
\eqno(2.3.4)
$$
Plugging that expression for $P$ inside the parent action (2.3.2) yields the action 
$$
S[Y^{ca\vert}{}_{b}] 
= \int d^{D}x\, \left( {1 \over 2} \,\partial_{c}Y^{ca\vert}{}_{b}\partial^{d}Y_{da\vert}{}^{b}
 - {1 \over 2(D-1)}\, \partial_{c}Y^{ca\vert}{}_{a}\partial^{b}Y_{bd\vert}{}^{d}\right)\;.
\eqno(2.3.5)
$$
\par
\medskip
 In order to analyse the gauge invariances of the action, 
it is sufficient to use only a decomposition of the 
various fields under $GL(D)$ and not under $O(1,D-1)\,$. 
Thus, we decompose 
$$
Y^{ab\vert}{}_{c} = X^{ab\vert}{}_{c} + \delta^{[a}_{c}\,Z^{b]}\;,\quad 
X^{ab\vert}{}_{a} \equiv 0\;.
\eqno(2.3.6)
$$
The invariance of the Maxwell action under the gauge transformation 
$\delta A_{a} = \partial_{a}\lambda$ is inherited by the new action (2.3.5), 
whereby the field 
$Z$ transforms as 
$\delta Z_{a} = \partial_{a}\lambda\,$, with $X^{ab\vert}{}_{c}$ staying unchanged, 
{\it i.e.} the action (2.3.5) can be shown to be invariant under
$$
\delta_{\lambda} Y^{ab\vert}{}_{c} = \delta^{[a}_{c}\partial^{b]}\lambda\;.
\eqno(2.3.7)
$$
On the other hand, from the fact that the field $Y^{ab\vert}{}_{c}$ enters the action 
(2.3.5) only through its divergence $\partial_{c}Y^{ca\vert}{}_{b}\,,$ the action is 
manifestly invariant under the following gauge transformation
$$
\delta_{\Upsilon}Y^{ab\vert}{}_{c} = \partial_{d}\Upsilon^{dab\vert}{}_{c}\;,\quad 
\Upsilon^{dab\vert}{}_{c} \equiv \Upsilon^{[dab]\vert}{}_{c}\;.
\eqno(2.3.8)
$$
Using the invariant antisymmetric symbol of $SL(D)\,$ to dualise the first two indices 
of $Y^{ab\vert}{}_{c}$, the decomposition (2.3.6) is tantamount to the following 
$GL(D)$-irreducible decomposition 
$$
\widetilde{Y}_{a[D-2]\vert c} = T_{a[D-2]\Vert c} + \widetilde{Z}_{a[D-2]c}\;, \quad 
T_{a[D-2] \Vert a}\equiv 0\;,\quad  \widetilde{Z}_{a[D-2]c} \equiv \widetilde{Z}_{[a[D-2]c]}\;,
$$
$$
\widetilde{Y}_{a[D-2]\Vert c} := {1\over 2} \epsilon_{a[D]} Y^{a[2]|}{}_{c}\;,\quad 
T_{a[D-2]\Vert c} := {1\over 2} \epsilon_{a[D]} X^{a[2]|}{}_{c}\;,\quad 
\widetilde{Z}_{a[D-2] c} := {1\over 2} \epsilon_{cba[D-2]} Z^{b}\;,
\eqno(2.3.9)
$$
while the gauge parameter $\Upsilon$ is similarly dualised into 
$$
\widetilde{\Upsilon}_{a[D-3]\vert c} = \lambda^{(1)}_{a[D-2]\Vert c} 
+ \lambda^{(2)}_{a[D-2]c}\;, \quad 
\lambda^{(1)}_{a[D-3] \Vert a}\equiv 0\;,\quad  \lambda^{(2)}_{a[D-3]c} \equiv 
\lambda^{(2)}_{[ a[D-3]c ]}\;.
\eqno(2.3.10)
$$
At this stage, without losing any of the tensorial fields involved, 
we set $D=4$ for the sake of clarity 
and to further explain the gauge structure of the new action in terms of the 
$GL(4)$-irreducible dual fields $T_{[2,1]}$ and $Z_{[1]}\,$.
The gauge transformations leaving the action (2.3.5) invariant, 
with $D=4$ and keeping the vector field $Z_{a}$
instead of its Hodge dual $\widetilde{Z}_{a[3]}\,$, now read
$$
\delta T_{ab\Vert c} = 2\,\partial_{[a}\lambda^{(1)}{}_{b]c}  - 2\,\partial_{[a}\lambda^{(2)}{}_{b]c}  
+ 2\,\partial_{c}\lambda^{(2)}{}_{ab}\;,
\eqno(2.3.11)
$$
$$
\delta Z_{a} = \partial_{a}\lambda + \partial^{b} \widetilde{\lambda}^{(2)}{}_{ab}\;,
\quad \widetilde{\lambda}^{(2)}{}_{ab} = {1\over 2}\epsilon_{abcd}{\lambda}^{(2)}{}^{cd}\;,
$$
where $\lambda^{(1)}{}_{ab}=\lambda^{(1)}{}_{(ab)}$ and 
$\lambda^{(2)}{}_{ab}=\lambda^{(2)}{}_{[ab]}\,$. 
\medskip 
In terms of the fields $X^{ab\vert}{}_{c}$ and $Z_{a}\,$ that we keep for the moment, 
the equations of motion derived from (2.3.5) are 
$$
0 = G_{ac\vert}{}^{b} := {1 \over 2}\,( \partial^{b}F_{ac}(Z)
+ 2 \,\partial^{d}\partial_{[c}X_{a]d\vert}{}^{b})\;, \quad F_{ac}(Z) := 2\,\partial_{[a}Z_{c]}\;.
\eqno(2.3.12)
$$
When the field $X$ is expressed in terms of its dual $T$, in four dimensions, we have 
the field equations
$$
\partial_{b}F_{ac}(Z) + {1 \over 12}\, \left[ 
\epsilon_{aduv}\partial_{c}\,F^{duv\Vert}{}_{b} - \epsilon_{cduv}\partial_{a}\,
F^{duv\Vert}{}_{b}  \right] = 0\;,\quad 
F_{abc\Vert d} := 3\,\partial_{[a}T_{bc]\Vert d}\;,
\eqno(2.3.13)
$$
where we note that the curvature $F_{abc\Vert d}$ of 
$T$ is invariant under the $\lambda^{(1)}$ gauge symmetry. 
Dualising on the indices $ac$ gives
$$
2\, \partial_{d}\widetilde{F}^{ab}(Z) -  \partial_{c}\,F^{abc\Vert}{}_{d}= 0\;,
\quad{\rm where}\quad  \widetilde{F}^{ab}(Z) := {1 \over 2}\,\epsilon^{abcd}F_{cd}(Z)\;. 
\eqno(2.3.14)
$$
Antisymmetrising the left-hand side of the equations of motion (2.3.13) in its free indices, 
one finds  
$$
\partial_{a}\,F^{abc\Vert}{}_{c} = 0\;.
\eqno(2.3.15)
$$
In this equation, only the $GL(4)$-irreducible field $T_{[2,1]}$ appears and all the 
symmetries in (2.3.11) are preserved. 
The above field equation is nothing but the equation 
$$
G^{(1)}_{a_{1}a_{2}c\Vert}{}^{a_{1}a_{2}} = 0 \;,
\eqno(2.3.16)
$$
presented in (2.2.8), in the case where $D=4\,$. 
We note that, using the Hodge decomposition whereby a differential $p$-form 
can be written as the sum of three terms, one d-exact, one $*d*$-exact and the last one harmonic:
$$
\omega_{[p]} = {\rm d} p_{[p-1]} + *\,{\rm d}* q_{[p+1]} + r_{[p]}\;,\quad 
\{ *{\rm d}* , {\rm d} \} r_{[p]}=0\;,
$$ 
the field $Z_{a}$ can be set to zero using the $\lambda$ and $\lambda^{(2)}\,$ gauge parameters, 
while its harmonic piece can be obtained by integrating equation (2.3.13), thereby expressing 
it in terms of the physical components of $T\,$. 
In the gauge where the closed and co-closed parts of $Z_{a}$ vanish, 
one cannot use any $\lambda^{(2)}$ gauge parameters anymore and 
the remaining action and field equations are only invariant under the $\lambda^{(1)}$ 
gauge symmetry. 

\medskip 
\medskip 
\medskip
{\bf {3 The three form in eleven dimensions }}
\medskip 
\medskip
The eleven dimensional supergravity theory as originally formulated contains the graviton 
and  the 
three form as its bosonic sector [30]. How to  formulate the eleven dimensional action with a six form was discussed in reference  [31]. 
The $E_{11}$ non-linear realisation in eleven dimensions includes the usual fields for the 
graviton and three form as well as the six form and a field which is the dual of the graviton,  
but in addition it contains an infinite number of  fields with blocks of height nine added, 
see equation (1.1). 
Among these fields are the   $h_{[9,9,\ldots,9,8,1]}\,$.  
In this section we will repeat the considerations of the sections two, but  for the three form.  
We will show ! how the alternative dual descriptions of the degrees of freedom usually encoded  in the 
three form  arise naturally within the unfolded formulation. 
We find the equations of motion of the theory when described 
by any of these dual gauge fields and we will find an infinite set of duality relations that are first 
order in space-time derivatives and encode the dynamics. 
As we will discuss in the Conclusions, these 
relations should be contained in the non-linear realisation based on~$E_{11}$. 
\medskip 
{\bf 3.1 The unfolded  representation of the three-form}
\medskip
In what follows we will construct the unfolded representation of the three-form,  
that is both $SO(1,10)$ and gauge invariant. 
We recall that this representation is indecomposable, but can be mapped via 
harmonic expansion to Wigner's irreducible unitary representation of $ISO(1,10)$ for 
the three form. 
It can be found following the unfolding procedure given in [19], see also 
[20,22] and references therein.
Unlike in the previous section where we presented the unfolded formulation of 
Maxwell's theory using standard tensor calculus, in this section we give a more 
formal and compact account of the unfolded 
representation using differential form calculus and stress its conceptual basis. 
\par
Wigner's unitary irreducible representation of the Poincar\'e group $ISO(1,10)$ 
corresponding to the free, dynamical three-form 
in eleven dimensions can be mapped to an unfolded module consisting of 
an infinite set ${\cal T}$ of $SO(1,10)$-irreducible tensors  
$$
{\cal T} = \left\{  F_{a[4]}\,, \; F_{a[4]\Vert b}\,,\; F_{a[4]\Vert b(2)}\,,\; F_{a[4]\Vert b(3)}\,,\; \ldots \right\}\; ,
\eqno(3.1.1)
$$
where the notation $F_{a[4]\Vert b(n)}$ indicates a tensor that is separately antisymmetric
in its four indices $\left\{ a_{1}, a_{2}, a_{3}, a_{4} \right\}$ and totally symmetric in its $n$ 
indices 
$\left\{ b_{1}, b_{2}, \ldots , b_{n} \right\}\,$. 
The 
$SO(1,10)$-irreducibility of the tensors $\left\{ F_{a[4]\Vert b(n)}  \  , n=0,1,\ldots \right\}$
means that, besides being Young-projected,  the tensors are traceless, 
viz.
$$
F_{a[4]\Vert  a b(n-1)} \equiv 0 \; ,\quad
\eta^{a_1 b_{1}} F_{a[4]\Vert b(n)}\equiv 0\;,
\eqno(3.1.2) 
$$   
where we  recall our convention that indices at the same position (covariant or 
contravariant) and with the same Latin label are implicitly 
symmetrised,  or antisymmetrised, according to the context. 
We note that the difference between an $GL(11)$ and an $SO(1,10)$-irreducible 
tensor is given by the tracelessness property, here the second identity of (3.1.2).  
\par 
In terms of Young tableau, the tensor $F_{a[4]\Vert b(n)}$ is represented by
$$
\vbox{\offinterlineskip\cleartabs
\def\hr{\vrule height .4pt width 2.5em}
\def\vr{\vrule height15pt depth 5pt}
\def\cc#1{\hfill#1\hfill}
\+ \hr&\hr&\hr&\hr&\cr
\+ \vr\cc{$a_1$}&\vr\cc{$b_1$}&\vr\cc{$\ldots$}&\vr\cc{$b_n$}&\vr&\cr
\+ \hr&\hr&\hr&\hr&\cr 
\+ \vr\cc{$a_2$}&\vr&\cr
\+ \hr&\cr
\+ \vr\cc{$a_3$}&\vr&\cr
\+\hr&\cr
\+ \vr\cc{$a_{4}$}&\vr&&\cr
\+ \hr& &&\cr} \;.
\eqno(3.1.3)
$$
The action of the Poincar\'e group on the infinite set of tensors in (3.1.1) is given by 
$$
P{}_{b}F_{a[4]} = F_{a[4]\Vert b}\,,\quad 
P_{b_{2}}F_{a[4]\Vert b_{1}} = F_{a[4]\Vert b_{1}b_{2}} \;,\quad
P_{b_{3}}F_{a[4]\Vert b_{1}b_{2}} = F_{a[4]\Vert b_{1}b_{2}b_{3}} \;,\;\;
\ldots
\eqno(3.1.4) 
$$ 
while the Lorentz generators $M_{ab}$ act diagonally in ${\cal T}$ by the usual action. 
Up to this stage, although we have talked of tensors, we have used  no notion of spacetime. 
\medskip
Introducing a  spacetime , 
the action of the translation generators of the Poincar\'e group on the representation can be explicitly 
realised by taking them to be differentiation  with respect to the space-time coordinates, that is, 
$P_a= \partial_a$, whereupon equations (3.1.4) take the form  
$$
\partial_{b}F_{a[4]} = F_{a[4]\Vert b}\;,
\eqno(3.1.5)
$$
$$
\partial_{b_{2}}F_{a[4]\Vert b_{1}} = F_{a[4]\Vert b_{1}b_{2}}\;,
\eqno(3.1.6)
$$
$$
\partial_{b_{3}}F_{a[4]\Vert b_{1}b_{2}} = F_{a[4]\Vert b_{1}b_{2}b_{3}}\;,
\eqno(3.1.7)
$$
$$
\vdots
$$
\par
\medskip

The infinite set of differential equations (3.1.5-7) can  be compactly written upon introducing
Grassmann odd (resp. even) vector oscillators $\theta^{a}$ (resp. $u^{a}$) and forming the 
master field 
$$
F(x;\theta,u) = \sum_{n=0}^{\infty} {{1}\over{4!n!}}\; F_{a[4]\Vert b(n)}(x)\, \theta^{a_{1}}\ldots\theta^{a_{4}}\,
u^{b_{1}}\ldots u^{b_{n}}\;.
\eqno(3.1.8)
$$
It is also advantageous to write everything in terms of differential forms, 
by using the total exterior derivative d $ = dx^{\mu}\partial_{\mu}\,$, 
taking 
the $F_{a[4]\Vert b(n)}$ to be zero forms and  introducing the one-form 
$$
h^{a} := dx^{\mu} \,\delta^{a}_{\mu} 
\eqno(3.1.9)
$$ 
for Minkowski spacetime in Cartesian coordinates. 
In this setting, the infinite set of differential equations (3.1.5-7)  given above can 
be written in the form  
$$
[ {\rm d} - i\, h^{a}\rho_{_{{\cal T}}} (P_{a}) ] F(x;\theta,u) = 0 \;,
\eqno(3.1.10)
$$
where the translation generators are now  represented on the master field as follows: 
$$
\rho_{_{{\cal T}}}(P_{a}) = (-i){{\partial} \over {\partial u^{a}}}\;.
\eqno(3.1.11)
$$

 Explicitly, equations (3.3.5)--(3.1.7) now read
$$
{\rm d}F_{a[4]} = h^{c}\, F_{a[4]\Vert c}\;,
\eqno(3.1.12)
$$
$$
{\rm d}F_{a[4]\Vert b} = h^{c}\, F_{a[4]\Vert bc}\;,
\eqno(3.1.13)
$$
$$
{\rm d} F_{a[4]\Vert b(2)} = h^{c}\, F_{a[4]\Vert b(2)c}\;.
\eqno(3.1.14)
$$
\par
\medskip

Taking into account the $GL(11)$ irreducibility conditions, given in equation (3.1.2),  of the tensor on the right-hand side of  equation (3.1.5), one derives the relation 
$$
\partial_{a} F_{a[4]} = 0 \;, 
\eqno(3.1.15)
$$
which is locally solved, as usual, by $F_{[4]} = {\rm d}A_{[3]}\,$, 
introducing a three-form potential and its four-form field strength
$$
A_{[3]} = {1 \over 6}\,h^{a_{1}}\wedge h^{a_{2}}\wedge h^{a_{3}}\;A_{a[3]}\,,
\quad 
F_{a[4]} := {1\over 24} h^{a_{1}} \wedge \ldots \wedge h^{a_{4}}\;F_{a[4]}\;.
\eqno(3.1.16)
$$
We are using the notation that a number in square brackets without being accompanied by a letter 
denotes the degree of the form that the field belongs to, that is, 
$A_{[3]}$ is a form of degree three.  
The zero-form tensor $F_{a[4]}$ are thus the components of the four-form 
field strength $ F_{[4]} = {\rm d}A_{[3]}$. As usual the gauge field  $A_{[3]}$ is defined up to the exterior derivative of a two-form potential, namely 
$$
A_{[3]} \sim A_{[3]} + {\rm d} \Lambda_{[2]}\;.
\eqno(3.1.17)
$$
On the other hand, recalling that $F_{a[4]\Vert b}$ is not only $GL(11)$ 
but also $SO(1,10)$ irreducible, as given in equation (3.1.2), 
one derives the equation  
$$
\partial^{a}F_{a[4]} = 0 \;, 
\eqno(3.1.18)
$$
which together with equation (3.1.16), is the field equation of a dynamical three-form.  
We also not that the other equations (3.1.13), (3.1.14) {\it etc.} can be solved one after 
the others. They express the tensors $F_{a[4]\Vert b(n)}$ as the higher gradients of the tensors 
$F_{a[4]\Vert b(m)}$ for $m<n$ and so in terms of the {\it on-shell} dynamical three-form  
$A_{[3]}$:
$$
F_{a[4]\Vert b(n)} = 4\partial_{b_{1}}\partial_{b_{2}}\ldots \partial_{b_{n}}\partial_{[a_{1}}A_{a_{2}a_{3}a_{4}]}\;. 
\eqno(3.1.19)
$$ 
We note  that the $SO(1,10)$ properties of $F_{a[4]\Vert b(n)}$ are ensured by the equations 
of motion of the three form and the fact that partial derivatives commute.  
\par
To summarise, the irreducible unitary representation of equation (3.1.1) contains components that are individually subject to $SO(1,D-1)$ irreducibility conditions and once we take the space-time translations to be 
realised by space-time differentiation these conditions imply the well known equation of motion for a three form. This is a purely algebraic way of encoding the field equations 
and Bianchi identities of a dynamical three-form, a characteristic of unfolded dynamics. 
 
The underlying algebraic structure, captured by (3.1.12)--(3.1.14) together with 
$dA_{[3]} $ $={1\over 24} h^{a_{1}} \wedge \ldots \wedge h^{a_{4}}\;F_{a[4]}$ and 
${\rm d}h^{a}= 0\,$, 
is known as a free differential algebra and makes sense on a base manifold of arbitrary dimension.
Its initial data is given by the gauge functions for $A_{[3]}$  and the vielbeins $h^{a}$ together
with the infinite set of constants provided by the zero-forms at a given point $p_{0}$ of the manifold.
In particular, in eleven dimensions, the infinite set of zero-forms in ${\cal T}$ at a point 
$p_{0}$ with Cartesian coordinates $x_{0}^{\mu}\,$, together with the   differential equations (3.1.10), give the necessary data that enables one to reconstruct
an on-shell, dynamical three-form around that point $p_{0}$ using the  Taylor expansion 
$$
A_{a[3]}(x) = A_{a[3]}(x_{0}) + \sum_{n=1}^{\infty} {1\over n!}\; (x-x_{0})^{b_{1}}\ldots 
(x-x_{0})^{b_{n}} \,F_{a[3]b\Vert b(n-1)}(x_{0}) \;. 
\eqno(3.1.20)
$$
\par
\medskip
We would like to make some comments on gauge fixing. 
In the  light-cone coordinates $x^{\mu} = ( x^{-},x^{+},x^{i})\,$ we can choose the Lorentz frame in 
which the momentum is $k_{\mu}=(k_{-},k_{+}=0,k_{i}=0)\,$.
Then at the point $p_{0}$, the components $A_{-jk}$ and $A_{-+j}$ can be 
set to zero by fixing the gauge in equation (3.1.17) 
using the gauge parameters $\lambda_{ij}\,$ and $\lambda_{+i}\,$. 
Furthermore, the components $A_{+ij}$ are gauge-invariant and zero on-shell as the field 
equation is given by $k_{-}A_{+ij}=0\,$.  
As a result  one finds that the three-form potential   has all its components vanishing 
except for the purely transverse ones, for which 
$$
A_{ijk}(x_{0}) = {1\over k_{-}} \, F_{- ijk}(x_{0})\;.
\eqno(3.1.21) 
$$ 
Consequently, all the derivatives of the three-form, when evaluated in momentum space 
and in the chosen Lorentz frame, 
are therefore given by all the powers of $k_{-}$ times the Fourier transform of 
$A_{ijk}(x)\,$ and they transform in the following representation
$$
\vbox{\offinterlineskip\cleartabs
\def\hr{\vrule height .4pt width 2.5em}
\def\vr{\vrule height15pt depth 5pt}
\def\cc#1{\hfill#1\hfill}
\+ \hr&\hr&\hr&\hr&\cr
\+ \vr\cc{$-$}&\vr\cc{$-$}&\vr\cc{$\ldots$}&\vr\cc{$-$}&\vr&\cr
\+ \hr&\hr&\hr&\hr&\cr 
\+ \vr\cc{$i$}&\vr&\cr
\+ \hr&\cr
\+ \vr\cc{$j$}&\vr&\cr
\+\hr&\cr
\+ \vr\cc{$k$}&\vr&&\cr
\+ \hr& &&\cr} \;.
\eqno(3.1.22)
$$
These  coincide with all the non-vanishing on-shell derivatives of the field strength. This 
discussion follows the general arguments given in references [18] 
(for related discussions, see [27])  
and it is  the equivalent, for the three-form, of the Petrov decomposition of a metric in 
Riemannian geometry. 
\par
We next note how the gauge-for-gauge transformations, $ \delta\lambda_{[2]} = {\rm d} C_{[1]}\,$, act. 
In our chosen Lorentz frame the only gauge transformations that have a non-trivial  gauge-for-gauge transformation are  $\lambda_{-i}\,$ and $\lambda_{-+}\,$. These are also   the only gauge parameters which we did not use so far. They are subject to transformations that  involve the 
components $C_{i}$ and $C_{+}$ and these can be used to set these gauge parameters to zero, that is set  $\lambda_{-i}= 0=\lambda_{-+}\,$. We note that the component $C_{-}$ can be set to zero by the gauge-for-gauge-for-gauge parameter. 
\par 
Another, alternative and Lorentz-covariant way of analysing the physical content of the equations 
consists in 
Taylor expanding the gauge (and higher reducibility) parameters, the three-form components 
as well as the field strength, all evaluated on-shell, and comparing all the coefficients of the various 
powers of $(x-x_{0})$ at the point $p_{0}\,$. One sees that the constants 
$A_{abc}(x_{0})$ can be set to zero by the constants 
$\partial_{[a}\lambda_{bc]}(x_{0})$ (the latter not being constrained by the reducibility 
transformations). Similarly, at first order in the derivatives of the three-form, the 
constants $\partial_{(a}A_{b)cd}(x_{0})$ can be set to zero by the constants 
$\partial_{a}\partial_{[b}\lambda_{cd]}+\partial_{b}\partial_{[a}\lambda_{cd]}\,$
whereas the constants $\partial_{[a}A_{bcd]}(x_{0})$ are identified (up to a constant, irrelevant  
factor) with the constants $F_{abcd}(x_{0})\,$, {\it etc}. The outcome of this procedure 
is that all the derivatives $\partial_{c_{n}}\ldots \partial_{c_{1}}A_{a[3]}(x_{0})$ 
of the three-form at the point $p_{0}$ are set equal to the on-shell derivatives 
$\partial_{(c_{n}}\ldots \partial_{c_{2}}F_{c_{1})a[3]}(x_{0})\,$, thereby explaining (3.1.20). 
This way of counting physical degrees of freedom on-shell 
is the one adopted in unfolded dynamics~[19].
\par
\medskip

It is well known that rather than describe the degrees of freedom by a three form one can use 
a 6-form potential and we now explain this from the unfolded viewpoint. We begin with the 
relation 
$$
F^{a[7]} := {1\over 4!} \epsilon^{a[7]b[4]}\,F_{b[4]}\;,
\eqno(3.1.23)
$$
and transfer the properties of the unfolded dynamics of the three form given  in equations 
(3.1.5-7) to corresponding equations for the six form. 
The first unfolded equation (3.1.5) transforms in the $[4,1]$-irrep of 
$SO(1,10)$ and the resulting divergenceless  property of $F_{a[4]\Vert b}$ implies that $F^{a[7]}$ is 
d-closed:
$$
0 = \partial^{a}F_{ac[3]} \quad \Leftrightarrow\quad \partial^{a}F^{a[7]} = 0\;,
\eqno(3.1.24)
$$
while the $GL(11)$-irreducibility of $F_{a[4]\Vert b}$, that is the 
Bianchi identity of $F_{a[4]}\,$,  implies that 
$F^{a[7]}$ is divergenceless:
$$
\partial_{a}F_{a[4]}\equiv 0 , \quad \Leftrightarrow \quad \partial_{b}F^{ba[6]}=0\;.
\eqno(3.1.25)
$$
By the usual Poincar\'e lemma, equation (3.1.24)  implies that $F^{a[7]}$ can locally be written as 
$$
F^{a[7]} = 7 \partial^{a}A^{a[6]}\;.
\eqno(3.1.26)
$$ 
Thus we find the  usual  exchange the equations of motion with the Bianchi identities in equations (3.1.24) and (3.1.25). 
\par
\medskip

We now define $F_{a[7]\Vert b}$ by 
$$
F_{a[7]\Vert b} := \partial_{b}F_{a[7]} \;.
\eqno(3.1.27)$$
By virtue of equations (3.1.24) and (3.1.25), $F_{a[7]\Vert b}$ 
is an irreducible $SO(1,10)$ tensor as it is 
$GL(11)$-irreducible ( $F_{a[7]\Vert a}=0$) and traceless 
($F_{a[6]b\Vert }{}^{b}=0$). 
Completing the unfolding of the dual linearised $6$-form yields  the following tower of tensors 
$$
\widetilde{\cal T} = \{ F_{a[7]\Vert b(n)}\;, \quad n=0,1,\ldots , \}\;.
\eqno(3.1.28)
$$
The action of the Poincar\'e generators $P_{c}$ on the tensors in $\widetilde{\cal T}$ is given by 
$$
P_{c}\,F_{a[7]\Vert b(n)} = F_{a[7]\Vert cb(n)}= \partial_{c}\,F_{a[7]\Vert b(n)}\;.
\eqno(3.1.29)
$$
It follows from (3.1.23) and the above conventions for the action of the Poincar\'e translations that 
the tensors in $\widetilde{\cal T}$ of equation (3.1.28) and those in ${\cal T}$ of equation (3.1.1) are 
related by 
$$ 
F_{a[7]\Vert b(n)} =  {1\over 4!} \epsilon^{a[7]c[4]}\, F_{c[4]\Vert b(n)}, \quad 
n = 1,2,\ldots
\eqno(3.1.30)$$
The tensors in $\widetilde{\cal T}$ are  traceless as result of the relation 
$$
F^{a[6]c\Vert }{}_{cb(n-1)} =  {1\over 4!} \epsilon^{a[6]cd[4]}\, F_{d[4]\Vert cb(n-1)}  = 0\;,
\eqno(3.1.31) 
$$
and are  $GL(11)$-irreducible  as a consequence of 
$$
\epsilon^{a[8]d[3]}F_{a[7]\Vert ab(n-1)} =  {1\over 4!} \epsilon^{a[8]d[3]} \epsilon_{a[7]c[4]}\,
F^{c[4]\Vert }{}_{ab(b-1)} =  7! F^{d[3]a\Vert }{}_{ab(n-1)}=0\;. 
\eqno(3.1.32) 
$$ 
Hence the tensors $\widetilde{\cal T}$  of equation (3.1.28) belong to the $SO(1,10)$ 
Young tableau
$$
\vbox{\offinterlineskip\cleartabs
\def\hr{\vrule height .4pt width 2.5em}
\def\vr{\vrule height15pt depth 5pt}
\def\cc#1{\hfill#1\hfill}
\+ \hr&\hr&\hr&\hr&\cr
\+ \vr\cc{$a_1$}&\vr\cc{$b_1$}&\vr\cc{$\ldots$}&\vr\cc{$b_n$}&\vr&\cr
\+ \hr&\hr&\hr&\hr&\cr 
\+ \vr\cc{$\vdots$}&\vr&\cr
\+ \hr&\cr
\+ \vr\cc{$a_6$}&\vr&\cr
\+\hr&\cr
\+ \vr\cc{$a_{7}$}&\vr&&\cr
\+ \hr& &&\cr} \;.
\eqno(3.1.33)
$$
We can collect the tensors $\widetilde{\cal T}$  into a single object 
$$
F(x;\theta, u) = \sum_{n=0}^{\infty} {{1}\over{7!n!}}\; F_{a[7]\Vert b(n)}(x)\, 
\theta^{a_{1}}\ldots\theta^{a_{7}}\,
u^{b_{1}}\ldots u^{b_{n}}\;,
\eqno(3.1.34)
$$
for which equation (3.1.29) takes the form 
$$
[ {\rm d} - i\, h^{a}\rho_{_{\widetilde{\cal T}}}(P_{a}) ] F(x;\theta,u) = 0 \;,\quad {\rm where }
\quad
\rho_{_{\widetilde{\cal T}}}(P_{a}) = (-i){{\partial} \over {\partial u^{a}}} = \rho_{_{{\cal T}}}(P_{a}) \;.
\eqno(3.1.35)
$$
\par 
Although action principles are usually part of the definition of an unfolded system, it is nevertheless 
instructive to consider a parent action from which one can find both the action for  the three gauge 
form and that for the  six form gauge field:
$$
S[A_{[3]},F_{[7]}] = \int ( {\rm d} A_{[3]}\wedge F_{[7]} - {1\over 8}\, F_{[7]}\wedge * F_{[7]})\;,
\eqno(3.1.36)
$$
where $F_{[7]}$  and $A_{[3]}$ are independent fields. Extremising it with respect to $A_{[3]}$ gives $ d F_{[7]}=0$ and so $F_{[7]}= {\rm d}A_{[6]}$;  
substituting this  back in $S[A_{[3]},F_{[7]}]$, gives the standard action 
$S[A_{[6]}]\propto \int {\rm d}A_{[6]}\wedge * {\rm d}A_{[6]}\,$. The equation of motion for $F_7$ gives $F_7 \propto   *dA_3$  and substituting back we find the standard action for the three form. 
\par 
Alternatively, one can  start from the Palatini formulation for the $3$-form,
$$
S[A_{[3]},F^{a[4]}] = \int {1\over 7!} \,\epsilon^{b[4]c[7]}\,h_{c_{1}}\wedge\ldots\wedge h_{c_{7}}
({\rm d}A_{[3]} + {1\over 8}h_{c_{1}}h_{c_{2}}h_{c_{3}}h_{c_{4}}F^{c[4]})
F_{b[4]},
\eqno(3.1.37)
$$
where $F^{a[4]}$ is a zero-form and is an independent field and we recall that $h_c$ is defined in equation (3.1.9). 
Defining 
$$
F_{[7]} := {1\over 7!} \,\epsilon^{b[4]c[7]}F_{b[4]}\,h_{c_{1}}\wedge\ldots\wedge h_{c_{7}} \;,
\eqno(3.1.38)
$$
the Palatini action (3.1.37)  becomes identical to the action (3.1.36). 
The latter action will be used in section 3.3 where we shall generalise the action principle given above
for Maxwell theory to the case of the three form in eleven dimension and in the frame-like formulation. 

\medskip 
{\bf 3.2 Further dualisation of the three form}
\medskip

It is well-known  that rather than express the dynamics of the bosonic non-gravitational degrees of 
freedom  of eleven dimensional supergravity by a  three-form gauge field  one can instead use a 
six-form gauge field $A_{[6]}\,$, whose curvature $F_{[7]}$, at the linearised level,  is just   the Hodge 
dual of $F_{[4]}\,$. As explained in the introduction,  the non-linear realisation of the Kac-Moody algebra 
$E_{11}$ leads not only to the usual fields of eleven dimensional supergravity as well as a six form and 
dual graviton field, but also to the   infinite set of fields of equation (1.1) which were proposed to be 
equivalent ways of  describing  the dynamics [6]. In this section we will show how the next field on the duality chain of equation (1.1), the gauge field $A_{[9,3]}$, arises and we give its 
linearised dynamics. The duality relation involving the fields in the gravity sector was sketched in 
reference  [14] and some indications that one might be able to do this for any massless particle were 
discuss in [24].  
\par
As we explained for Maxwell theory in the previous section,  one can dualise any  of 
the curvature tensors that occur in the unfolded formulation.
Hence,  instead of dualising the first tensor in the set ${\cal T}$ in (3.1.1), 
one may dualise the second tensor $F_{[4,1]}$ on its second column:
$$
G^{b[10]}{}_{\Vert a[4]} = \epsilon^{b[10]c}\,F_{a[4]\Vert c}\;,
\eqno(3.2.1)
$$
or equivalently 
$$
F_{a[4]\Vert b}= -{1\over 10 !} \epsilon _{bc[10]} G^{c[10]}{}_{\Vert a[4]}\;.
\eqno(3.2.2)
$$
We can now find what the constraints on $F_{a[4]\Vert b}$ imply for 
$G_{b[10]}{}_{\Vert a[4]}$. Taking the trace of (3.2.2) and using the 
second equation in (3.1.2) we find that indeed, 
$$
G_{b[10]\Vert b a[3]} = 0\;,
\eqno(3.2.3)
$$
while using the first equation in (3.1.2) and acting with 
$\epsilon^{a[4]b d[6]}$ on equation (3.2.2) we find the quartic trace constraint
$$
G^{b[6]a[4]}{}_{\Vert  a[4]}\equiv ({\rm Tr}_{12}){}^{4} G_{[10,4]} = 0\;.
\eqno(3.2.4)
$$
The presence of a higher order trace condition is unusual when compared to the standard 
formulation of particle dynamics including Fronsdal's higher-spin dynamics. 
\par
Equation (3.2.3) implies that  the tensor $G_{a[10]\Vert b[4]}$ is an irreducible $GL(11)$ 
tensor of type $[10|4]$, but it is not an $SO(1,D-1)$-irreducible tensor as it does not satisfy a single 
trace condition. We note that, as usual, the  Bianchi identities and field equations get 
swopped  under the dualisation. 
\par
We would now like to look at the differential constraints on $G_{b[10]\Vert  a[4]}$ that arise 
from the differential constraints on $F_{a[4]\Vert c}$ of equations (3.1.5-7). The first of 
these equations implies that 
$\partial{}_{a} F_{a[4]\Vert b} = 0$ which using equation (3.2.2)  in turn implies that 
$$
\partial_{a}G^{b[10]}{}_{\Vert a[4]} = 0\;.
\eqno(3.2.5)
$$
As we did for the Maxwell case we can continue taking more space-time derivatives 
of the the field strength $G_{b[10]}{}_{\Vert a[4]}$ to find an infinite set of tensors 
$\{ G_{a[10] \Vert b[4]\Vert  b(n)},  \ n=0,1,\ldots\}\,$. Using similar arguments we can 
transfer  the properties of $F_{a[4]\Vert  b(n)} $ to those new tensors to find that 
$\{ G_{a[10] \vert b[4]\Vert  b(n)},  \ n=0,1,\ldots\}\,$ are $GL(11)$-irreducible 
and obey the trace  constraints 
$$
{(\rm Tr_{12})}^{4} G_{[10,4,1,\ldots,1]} = 0 \;,\qquad 
{\rm Tr_{1i}}G_{[10,4,1,\ldots,1]} = 0 = {\rm Tr_{2i}}G_{[10,4,1,\ldots,1]}\;,\quad 
i\in \{ 3,\ldots,n\}\;.
\eqno(3.2.6)
$$
The notation $({\rm Tr}_{ij})^n$ used here means that one takes $n$ traces on the columns 
$i$ and $j\,$.  
\par
Equation (3.2.5),  combined with the $GL(11)$ irreducibility of $G^{b[10]}{}_{\Vert a[4]}$ 
implies, using  the generalised Poincar\'e lemma [28], 
that it can locally be written as 
$$
G^{b[10]}{}_{\Vert a[4]} = \partial_{a}\partial^{b}A^{b[9]}{}_{\Vert a[3]}\;, 
\eqno(3.2.7)
$$ 
where the $GL(11)$-irreducible tensor gauge field $A_{b[9]}{}_{\Vert a[3]}$ is defined up to 
the gauge transformation
$$
\delta A_{a[9]\Vert b[3]} = 9\, \partial_{a}\Lambda^{(1)}{}_{a[8]\Vert b[3]} 
+ 3\, (\partial_{b} \Lambda^{(2)}{}_{a[9]\Vert b[2]} 
+ {9 \over 7} \,\partial_{a}\Lambda^{(2)}{}_{a[8]b\Vert b[2]})\;, 
\eqno(3.2.8)
$$
with the two gauge parameters being $GL(11)$-irreducible with type 
$\Lambda^{(1)}{}_{[8,3]}$ and $\Lambda^{(2)}{}_{[9,2]}\,$. 
We note that there are no algebraic trace constraints on $A_{[9,3]}$, nor on its gauge 
parameters. 
\par
Remembering the expression  $F_{a[4]\Vert b} = \partial_{b} \partial_{a}A_{a[3]}\,$, the definition (3.2.1) of $G_{a[10]\Vert b[4]}$ and the relation (3.2.7)  give us the following 
duality relation:
$$
\partial^{a}A^{a[9]\Vert}{}_{b[3]} = \epsilon^{a[10]c}\,\partial_{c} A_{b[3]} 
+ \partial_{b}\Xi^{a[10]\vert}{}_{b[2]}\;.
\eqno(3.2.9)
$$ 
which is the analog of (2.2.38).
We first note that $\Xi^{a[10]\vert}{}_{b[2]}$ decomposes into 
$$
\Xi^{a[10]\vert}{}_{b[2]} = \Xi^{(1)}{}^{a[10]}{}_{\Vert b[2]} + 
\epsilon^{a[10]}{}_{b} \Xi^{(2)}{}_{b}
\eqno(3.2.10)
$$
and that a gauge transformation $A_{b[3]} \rightarrow A_{b[3]} + \partial_{b}\lambda_{b[2]}$
with $\lambda_{b[2]}= - x_{b}\Xi^{(2)}{}_{b}$ enables one to eliminate the $\Xi^{(2)}$
component of $\Lambda\,$. 
Having done that, the equation (3.2.9) is now understood with a field 
$\Xi_{a[10]\Vert b[2]}$ obeying $\Xi_{a[10]\Vert ab}=0\,$. 
We can now reformulate this equation in the same manner as we did for equation (2.2.38). 
By shifting the arbitrary field
$\Xi^{a[10]\Vert}{}_{b[2]}$ in an appropriate way,  
we can recast the equation in the form 
$$
\partial^{a}A^{a[9]\Vert}{}_{b[3]} = \epsilon^{a[10]b}\,4\partial_{b} A_{b[3]]} 
+ \partial_{b}\Xi^{a[10]\Vert}{}_{b[2]}\;.
\eqno(3.2.11)
$$
Multiplying by $\epsilon^{a[10]e}$ and tracing on $b_1$ and $e$,  we find that 
$\partial ^a \Xi^{a[10]\Vert}{}_{b[2]}=0 $  implying that 
$ \Xi^{a[10]\Vert}{}_{b[2]}= \partial ^a\Xi^{a[9]\Vert}{}_{b[2]}$. 
Using this result  equation (3.2.9) now  becomes 
$$
\partial^{a}A^{a[9]\Vert}{}_{b[3]} = \epsilon^{a[10]c}\,4\partial_{[c} A_{b[3]]} 
+ \partial_{b}\partial^a\Xi^{a[9]\Vert}{}_{b[2]}\;.
\eqno(3.2.12)
$$ 
We recognise the last term as a gauge transformation of the field  
$A^{a[9]\Vert}{}_{b[3]}$. Alternatively, the above equation can be made fully gauge 
invariant by giving a shift symmetry to the field $\Xi\,$ under $\Lambda^{(2)}{}_{[9,2]}$.
\par 
\medskip

We now give an action principle for the $A_{b[9]}{}_{\Vert a[3]}$ potential that
correctly describes the degrees of freedom of a massless three-form. 
The procedure was proposed in [14], which itself was inspired from 
[6,9]. 
We start with the three-form and follow the analog of the procedure for the Maxwell field 
spelled out in Section 2.3. 
To this end, we take the usual action $S[A_{[3]}]$ for a three form and integrate by parts, ignoring 
boundary terms:
$$
-{1\over 4!}\int d^{D}x\;\partial_{a}A_{a[3]}\partial^{a}A^{a[3]} = 
-{1\over 3!}\int d^{D}x\;\left( 
\partial_{b}A_{a[3]}\partial^{b}A^{a[3]} + 3 \partial^{b}A_{ba[2]}\partial_{c}A^{ca[2]} 
\right)\;.
\eqno(3.2.13) 
$$
\noindent 
We then introduce the following parent action, that features two independent fields, 
$P_{b|a[3]}$ and $Y^{b[2]|a[3]}$:
$$
S[P,Y] = -{1\over 3!}\int d^{D}x\;\left( P_{b|a[3]}\partial_{c}Y^{cb|a[3]} + 
P_{b|a[3]}P^{b|a[3]} + 3 P^{b|}{}_{ba[2]}P_{c|}{}^{ca[2]}\right) \;.
\eqno(3.2.14)
$$
Varying the action $S[P,Y]$ with respect to the field $Y^{a[2]|b[3]}$ gives the equation 
$\partial _{b_1} P_{b_{2}|a[3]} = 0$ which implies that $P_{b|a[3]} = \partial_{b}A_{a[3]}$. 
Substituted inside the action, we reproduce the action (3.2.13). 
On the other hand, as the field $P_{b|a[3]}$ is auxiliary one can 
express it in terms of $Y$ via its equation of motion, namely 
$$
2 P^{b|a[3]} = -\partial_{c}Y^{cb|a[3]} - {3\over D-1}\,\eta^{ba} \partial_{c}Y^{cd|}{}_{d}{}^{a[2]}\;,
\eqno(3.2.15)
$$
and substitute for it into the parent action, thereby 
yielding a daughter action $S[Y^{b[2]|a[3]}]$ expressed solely in terms of the field $Y\,$: 
$$
S[Y^{cb|a[3]}] = {1\over 4!}\int d^{D}x\;\Big( P_{b|a[3]}\partial_{c}Y^{cb|a[3]} + 
\partial_{c}Y^{cb|a[3]}\partial^{e}Y_{eb|a[3]} 
$$
$$
- { 3(5D^{2}-11D+7) \over (D-1)^{2}} \;
\partial_{c}Y^{cb|}{}_{ba[2]}\,\partial^{e}Y_{ed|}{}^{da[2]} \Big) \;.
\eqno(3.2.16)
$$
Setting $D=11\,$, one can then dualise $Y^{a[2]}{}_{|b[3]}$ on its first two indices, 
and decompose
$$
\widetilde{Y}_{a[9]\vert b[3]} = {1 \over 2}\epsilon_{a[9]c[2]}\, Y^{c[2]}{}_{|b[3]} = 
A_{a[9]\Vert b[3]} + B_{a[9] b\Vert b[2]} + \epsilon_{a[9]b[2]}C_{b}\;,
\eqno(3.2.17)
$$
so as to produce the $GL(11)$-irreducible field $A_{a[9]\Vert b[3]}\,$ 
satisfying $A_{a[9]\Vert ab[2]}\equiv 0\,$, as well as 
$B_{a[10] \Vert b[2]}$ (satisfying $B_{a[10]\Vert ab}\equiv 0\,$) and $C_{a}\,$
which are analogs of the field $\widetilde{Z}$ in Equation (2.3.9).
Because the field $Y^{cb|a[3]}$ enters the action only through its divergence 
$\partial_{c}Y^{cb|a[3]}$, the action is invariant under the following gauge 
transformations
$$
\delta Y^{b[2]|a[3]} = \partial_{c}\Upsilon^{c[3]\vert a[3]}\;,
\eqno(3.2.18) 
$$
where the gauge parameter $\Upsilon$ is antisymmetric in its two groups of indices. 
Upon dualising the parameter $\Upsilon\,$, one gets the following $GL(11)\,$-irreducible gauge 
parameters 
$$
{1 \over 3! }\,\epsilon_{c[3]} \Upsilon^{c[3]\vert a[3]} \longrightarrow 
\{ \Lambda^{(1)}_{a[8]\Vert b[3]}\;, \Lambda^{(2)}_{a[9]\Vert b[2]}\;, 
\Lambda^{(3)}_{a[10]\Vert b}\;, \Lambda^{(4)} \}\;. 
\eqno(3.2.19)
$$
The field $A_{a[9]\Vert b[3]}\,$ will then transform as in (3.2.8), while the gauge 
transformation of the field 
$B_{a[10] \Vert b[2]}$ will involve the gradient of the 
parameters $\Lambda^{(2)}$ and $\Lambda^{(3)}\,$. 
Finally, the vector field $C_{a}$ will transform with the gradient of $\Lambda^{(4)}\;$. 
\par
We note that the action also possesses the gauge symmetry involving the two-form gauge parameter 
$\lambda_{a[2]}$ inherited from the original three-form $A_{a[3]}\,$. 
This will be discussed in the next section 3.3,
where we use the frame-like formalism that brings in a better insight into the gauge structure.
On-shell, the gauge field $A_{a[9]\Vert b[3]}\,$ will obey the equation (3.2.4) discussed above.

\medskip 
{\bf 3.3 Unfolded description containing the $A_{[9,3]}$ form}
\medskip

In this section we wish to construct the unfolded formulation of the dynamics for 
the $A_{[9,3]}$ form, that is a set 
of first order differential equations that contain the gauge field $A_{a[9]\Vert b[3]}\,$ and that 
assumes the form of a free differential algebra. 
This will contain the manifestly Lorentz covariant and gauge-invariant infinite-dimensional 
representation of $ISO(1,D-1)$ constructed from the  field 
strength,  $G_{a[10] \Vert b[4]}\,$, and  all of its higher on-shell derivatives. It also contains the gauge field $A_{a[9]\Vert b[3]}$ through an appropriate frame-like, or 
Cartan-like, connection. In the next subsection 3.4, we will build an action principle for the $A_{a[9]\Vert b[3]}\,$
potential, but this time facilitated by the use of the frame-like description 
that we first derive on-shell in the present subsection.  
\par
\medskip

We first introduce, following  [20], the connection-like objects  
$$
\{ e_{[9]}{}^{a[3]}, \;\omega_{[3]}{}^{a[10]}\, \} \;.
\eqno(3.3.1)
$$
The indices in square brackets without a label, i.e. $[3]$ and $[9]$,  denote the form degree 
of the objects, for example $e_{[9]}{}^{a[3]}$ is a nine form that carries three 
antisymmetrised tangent indices and so can be written in more usual notation as 
${1 \over 9!}\,h^{b_1}\wedge \ldots \wedge h^{b_9} e_{{b_1}\ldots b_9} {}^{a_1a_2a_3}\,$. 
The field $\omega_{[3]}{}^{a[10]}$ is a three form that carries ten antisymmetrised 
tangent indices.  
It is important to note that the objects of equation (3.3.1) are not subject to  any  
$GL(D)$ irreducibility conditions. 
By analogy with the vielbein formulation of general relativity, we may think of 
$e_{[9]}{}^{a[3]}$ as a generalised vielbein and $\omega_{[3]}{}^{a[10]}$ as a generalised 
spin-connection. 
As the field $e_{[9]}{}^{a[3]}$ is not $GL(D)$ irreducible, only one of its irreducible 
components can be identified with  the gauge potential $A_{b[9] \Vert a[3]}$ that we 
considered in section 3.2; the precise  identification will be discussed below.  
\par
The differential forms of equation (3.3.1) are required to satisfy the differential equations 
$$
{\rm d}e_{[9]}{}^{a[3]} + h^{b_{1}}\wedge \ldots \wedge h^{b_{7}}\wedge
\omega_{[3]}{}^{a[3]}{}_{b[7]} = 0\;,
\eqno(3.3.2)
$$
$$
{\rm d} \omega_{[3]} {}^{a[10]} + h^{c_{1}}\wedge \ldots \wedge h^{c_{4}}\; G^{a[10]\Vert}{}_{c[4]} = 0\;,
\eqno(3.3.3)
$$
 where by assumption $G^{a[10]\Vert}{}_{c[4]}$ is the zero-form that appeared in (3.2.1). 
It obeys the $GL(D)$ irreducibility conditions and 
the higher-trace constraints of equations (3.2.3), (3.2.4).  
As discussed (3.2.5), this zero-form is the first member of an infinite set of zero-forms obeying the 
following first-order differential constraints:
$$
{\rm d}G^{a[10]\Vert b[4]} + h_{c } G^{a[10]\Vert b[4]\Vert  c} = 0\;,
\eqno(3.3.4)
$$
$$
{\rm d}G^{a[10]\Vert b[4]\Vert  c} + h_{c } G^{a[10]\Vert b[4]\Vert  c(2)} = 0\;,
\eqno(3.3.5)
$$
$$
\vdots
$$
$$
{\rm d}G^{a[10]\Vert b[4]\Vert  c(n)} + h_{c } G^{a[10]\Vert b[4]\Vert  c(n+1)} = 0\;,\quad n=2,3,\ldots \;.
\eqno(3.3.6)
$$
The equations (3.3.2)--(3.3.6) together with $dh^{a}=0$ form a free differential algebra and 
provides the unfolded description of the dual $A_{[9,3]}$ metric-like gauge field. 

The gauge transformations of the system (3.3.2)-(3.3.3) are
$$
\delta_{\epsilon} e_{[9]}{}^{a[3]} = {\rm d} \epsilon_{[8]}{}^{a[3]}
+ h_{a_{1}}\wedge \ldots \wedge h_{a_{7}}\wedge \epsilon_{[2]}{}^{a[10]} = 0\;,
\eqno(3.3.7)
$$
$$
\delta_{\epsilon} \omega_{[3]}{}^{a[10]}  = {\rm d}\epsilon_{[2]}{}^{a[10]}\;.
\eqno(3.3.8)
$$
The algebraic, St\"uckelberg-like, gauge transformations on  $ e_{[9]}{}^{a[3]}$, that is 
those contained in $\epsilon_{[2]}{}^{a[10]}\,$, can be used to  gauge away certain 
components of $ e_{[9]}{}^{a[3]}$. 
The $GL(11)$-irreducible decompositions of $ e_{[9]}{}^{a[3]}$
and $\epsilon_{[2]}{}^{a[10]}\,$ are respectively given by 
$$
[9]\otimes [3]\cong [11,1] \oplus [10,2]\oplus [9,3]\;, \quad {\rm and}
\eqno(3.3.9)
$$
$$
[10]\otimes [2] \cong [11,1] \oplus [10,2]\;.
\eqno(3.3.10)
$$
Therefore, after using all the algebraic gauge symmetries, the remaining components in 
$ e_{[9]}{}^{a[3]}$ are contained in the $GL(11)$-irreducible gauge field 
$A_{a[9] \Vert b[3] }\,$. Thus we make the connection with the equations of motion of 
section 3.2 which involved the $GL(D)$-irreducible gauge field $A_{a[9] \Vert b[3] }\,$. 
The connection $\omega_{[3]}{}^{a[10]} $ possesses two $GL(11)$-irreducible pieces: 
$[10,3]\oplus[11,2]\,$. However, it is determined from the ``zero-torsion'' equation (3.3.2) 
by the first derivatives of the components of ${\epsilon}_{[9]}{}^{a[3]}\,$ that can 
be reduced (or gauge-fixed) to its $A_{a[9] \Vert b[3] }$ part. 
As a result we find that only the $[10,3]$ irreducible component of 
$\omega_{[3]}{}^{a[10]} $ remains that we denote by $\tilde \omega _{a[10] \Vert b[3]}$. 
\par
In summary, so far, Equations (3.3.4)--(3.3.6) constrain a tower of manifestly Lorentz-covariant and 
gauge-invariant zero forms $\{ G^{(n)}\;, n=0,1,\ldots \}$ such that these can be expanded 
in terms of the unitary and irreducible massless representation of $ISO(1,D-1)$  that 
describes the degrees of freedom propagated by the original three form gauge field.
This is simply a consequence of the fact that the field strength $G^{a[10]\Vert b[4]}$ is by assumption expressed in terms of the field strength $F_{a[4]}$ 
via (3.2.1), so the representation appearing in (3.3.4)-(3.3.6) is equivalent to the 
representation built on the field strength $F_{a[4]}$ contained in equation (3.1.1).  
Equations (3.3.2) and (3.3.3) glue the zero-form tower to the 
gauge field $e_{[9]}{}^{a[3]}$ thanks to the introduction of the generalised spin connection 
$\omega_{[3]}{}^{a[10]}$ so as to write the full system as a free differential algebra. 
 We will show later in this section how to reproduce an equivalent
dynamics from an action principle involving the fields in (3.3.1) with some additional zero-forms. 
\par
In order to make contact with the gauge parameters of the metric-like $A_{[9,3]}$ gauge fields, 
we note that, as is typical for $p$-form systems such as a nine-form and a 3-form, the gauge 
transformations admit reducibility transformations. 
The complete family of gauge-for-gauge $p$-form parameters, which are not 
$GL(D)$ irreducible,  is given by:
$$
\{ {\epsilon}_{[9-i]}{}^{a[3]}  \} \;,\quad i=1,2,\ldots, 9
\eqno(3.3.11)
$$
and 
$$
\{ {\epsilon}_{[3-j]}{}^{a[10]}  \} \;,\quad j=1,2,3
\eqno(3.3.12)
$$
with transformation rules
$$
\delta {\epsilon}^{\,a[3]}_{[8]} = {\rm d} {\epsilon}^{a[3]}_{[7]}
+ h_{a_{1}}\wedge \ldots \wedge h_{a_{7}}\wedge \epsilon_{[1]}^{a[10]}\;,
\quad \delta_{\epsilon} {\epsilon}^{\,a[3]}_{[7]} = {\rm d} {\epsilon}^{a[3]}_{[6]}
+ h_{a_{1}}\wedge \ldots \wedge h_{a_{7}}\wedge \epsilon_{[0]}^{a[10]}\;,
\eqno(3.3.13)$$ 
$$
\delta \epsilon^{a[10]}_{[2]} = {\rm d} {\epsilon}^{\,a[10]}_{[1]}\;, \quad 
\delta \epsilon^{a[10]}_{[1]} = {\rm d} {\epsilon}^{\,a[10]}_{[0]}\;, \quad
\delta \epsilon^{a[10]}_{[0]} = 0\;,
\eqno(3.3.14)$$
and 
$$
\delta {\epsilon}^{\,a[3]}_{[6]} = {\rm d} {\epsilon}^{\,a[3]}_{[5]}\;, 
\quad  
\delta {\epsilon}^{\,a[3]}_{[5]} = {\rm d} {\epsilon}^{\,a[3]}_{[4]}\;, 
\ldots \;,\quad
\delta {\epsilon}^{\,a[3]}_{[1]} = {\rm d} {\epsilon}^{\,a[3]}_{[0]} \;, 
\quad \delta {\epsilon}^{\,a[3]}_{[0]} = 0\;.
\eqno(3.3.15)
$$
The  gauge-for-gauge parameter $\epsilon^{a[10]}_{[1]}$ can be used to gauge away parts of 
the parameter ${\epsilon}^{\,a[3]}_{[8]}$. 
Both are $GL(11)$ reducible and can be decomposed into the  $GL(11)$ representations as follows 
$$
[10]\otimes [1]\cong [11] \oplus [10,1] ,\quad 
[8]\otimes [3]\cong [11] \oplus [10,1] \oplus [9,2]\oplus [8,3]
\eqno(3.3.16)
$$
As this decomposition makes clear we can gauge away two components leaving the gauge parameter 
$\epsilon^{a[3]}_{[8]}\,$ to contain only the 
$GL(11)$-irreducible representation $[9,2]\oplus[8,3]$. Making the appropriate $GL(11)$ 
projection on equation (3.3.7),  we find that the 
gauge transformation of the $A_{[9,3]}$ potential takes the form :
$$
\delta_{\epsilon}A_{a[9]\Vert }{}^{b[3]} = 9\,\partial _{a} \epsilon_{a[8]\Vert}{}^{b[3]}
+ 3\,(\partial^{b} \epsilon_{a[9]\Vert}{}^{\ b[2]} + {9 \over 7}\,\partial^{b} 
\epsilon_{a[8]}{}^{b\Vert b}{}_{a} )\;,
\eqno(3.3.17)
$$
 thereby making contact with (3.2.8).
\par
When equations (3.3.2)-(3.3.3) are reduced to the remaining $GL(11)$-irreducible components 
$A_{a[9]\Vert b[3]}$ and  
$\tilde \omega_{a[10]\Vert b[3]}$ of $e_{[9]}{}^{a[3]}$ and $\omega_{[3]}{}^{a[10]}$, they
 become  
$$
\tilde \omega_{a[10]\Vert b[3]} = \partial_{a} A_{a[9] \Vert b[3]}\;,
\eqno(3.3.18)$$ 
$$
\partial^b \tilde \omega_{a[10]\Vert}{}^{b[3]} = G_{a[10]\Vert }{}^{b[4]}\;.
\eqno(3.3.19)
$$ 
The expression of the field strength in terms of the gauge field is given by 
$$
\partial^{b} \partial_{a} A_{a[9]\Vert }{}^{b[3]} = G_{a[10]\Vert}{}^{b[4]}\;,
\eqno(3.3.20)
$$ 
which agrees with equation  (3.2.7). 
It is easy to see that it is invariant under the gauge transformations (3.3.17). 
\par
\medskip

\medskip 
{\bf 3.4 First-order frame-like action for the $A_{[9,3]}$ field}
\medskip

We now follow the general procedure explained in [24], whose discussion for the   spin-2 case was already given in  [29].  
The action, just like the one given at the end of section 3.3, is a parent 
action in the sense that it contains both the three form and the $A_{[9,3]}$ gauge field. 
The difference between this action and the one presented in section 3.3 is that we will now use the 
frame-like vantage point developed above  for the gauge field 
$A_{a[9]\Vert b[3]}\,$.
We start from the action principle for the three-form, 
written in the Palatini formulation presented at the end of section 3.1 and that we repeat here for 
convenience: 
$$
S[A_{[3]},F^{a[4]}] = \int_{{\cal M}_{11}} {1\over 7!} \,\epsilon^{b[4]c[7]}\,h_{c_{1}}\wedge\ldots\wedge h_{c_{7}}
({\rm d}A_{[3]} + {1\over 8}h_{c_{1}}h_{c_{2}}h_{c_{3}}h_{c_{4}}F^{c[4]})
F_{b[4]},
\eqno(3.4.1)
$$
where $A_{[3]}$ and $F_{a[4]}$ are independent fields. 
We next introduce the parent action 
$$
S^{P}[A_{[3]},F_{a[4]},t_{[1]}{}^{a[3]},e_{[9]}{}^{a[3]}] = \int_{{\cal M}_{11}}  
\Big[\;{1\over 7!} \,\epsilon^{b[4]c[7]}\,h_{c_{1}}\wedge\ldots\wedge h_{c_{7}}\qquad \qquad
\qquad \qquad\qquad \qquad
$$
$$
\qquad \qquad \wedge\Big({\rm d}A_{[3]} + {1\over 8}h_{c_{1}}h_{c_{2}}h_{c_{3}}h_{c_{4}}F^{c[4]} 
+ t_{[1]}{}^{c[3]}h_{c_{1}}h_{c_{2}}h_{c_{3}} \Big) 
F_{b[4]}  \;+\;  t_{[1]\,a[3]}\,{\rm d}e_{[9]}{}^{a[3]}  \Big] \;,
\eqno(3.4.2)
$$
that contains the additional independent fields $t_{[1]\,a[3]}$ and $e_{[9]}{}^{a[3]}$.
The field equations derived from the parent action are given by 
$$
{\rm d}A_{[3]} + {1\over 4}h_{c_{1}}h_{c_{2}}h_{c_{3}}h_{c_{4}}F^{c[4]} 
+ t_{[1]}{}^{c[3]}h_{c_{1}}h_{c_{2}}h_{c_{3}} = 0\;,
\eqno(3.4.3)
$$
$$
{\rm d} * (h^{a_{1}}\ldots h^{a_{4}}\;F_{a[4]}) = 0\;,
\eqno(3.4.4)
$$
$$
{\rm d}\, t_{[1]}{}^{a[3]} = 0\;,
\eqno(3.4.5)
$$
$$
{\rm d} e_{[9]}^{a[3]} +{1\over 7!} \, h^{a_{1}}h^{a_{2}}h^{a_{3}}\,F_{b[4]}\,
\epsilon^{b[4]c[7]}\,h_{c_{1}}\wedge\ldots\wedge h_{c_{7}} = 0\;.
\eqno(3.4.6)
$$
The gauge symmetries of the action are 
$$
\delta A_{[3]}  = {\rm d} \lambda_{[2]}  - h_{c_{1}}h_{c_{2}}h_{c_{3}}\;\psi_{[0]}{}^{c[3]}\;,
\eqno(3.4.7)
$$
$$
\delta F_{[0]}{}^{a[4]} = 0\;,
\eqno(3.4.8)
$$
$$
{\delta} t_{[1]}{}^{a[3]} = {\rm d}\psi_{[0]}{}^{a[3]}\;,
\eqno(3.4.9)
$$
$$
\delta e_{[9]}{}^{a[3]}  = {\rm d} \xi_{[8]}{}^{a[3]} \;.
\eqno(3.4.10)
$$
The equation (3.4.5) results from extremising with respect to $e_{[9]}{}^{a[3]}$. 
It implies that $t_{[1]}^{a[3]}= {\rm d} C^{a[3]}$ and substituting this into the action we 
can absorb $ C^{a[3]}$ into $A_{[3]}$ and the action becomes that of equation 
(3.4.1). 
\medskip

In order to descend on a different child action, we note that by using the St\"uckelberg-like gauge symmetry of $A_{[3]}$ with gauge parameter
$\psi_{[0]}{}^{a[3]}\,$, one can completely gauge $A_{[3]}$ away, so that it disappears 
from the action (3.4.2). One can still perform differential gauge transformations 
with the two-form gauge parameter $\lambda_{[2]}\,$, but in order to stay in the 
gauge where $A_{[3]}$ is zero, one has to compensate it with a residual 
transformation with parameter 
$\bar\psi^{a[3]} = \partial^{[a_{1}}\lambda^{a_{2}a_{3}]}\,$. 
\par
\medskip

We are thus left with a child action containing the fields, 
$e_{[9]}{}^{a[3]}\,$, $F_{a[4]}\,$ and $t_{[1]}{}^{a[3]}\,$, 
or better, containing  only $e_{[9]}{}^{a[3]}\,$ and $t_{[1]}{}^{a[3]}\,$ as 
$F_{a[4]}$ can be expressed in terms of the totally antisymmetric part of 
$t_{[1]}{}^{a[3]}$ via (3.4.2). Note that  
under ${\delta} t_{[1]}{}^{a[3]} = {\rm d}\psi_{[0]}{}^{c[3]}\,$ and in the gauge where
$A_{[3]}=0\,$, we have that 
$\psi^{a[3]}=\bar\psi^{a[3]}=\partial^{[a_{1}}\lambda^{a_{2}a_{3}]}$ and therefore 
${\delta} t_{[b\vert a_{1}a_{2}a_{3}]} = 0\,$, as it should. 
The field $t_{[1]}{}^{a[3]}\,$ plays the role of the 
connection $\omega_{[3]}{}^{a[10]}\,$ introduced in (3.3.1), upon dualisation 
of $t_{[1]}{}^{a[3]}\,$ on its form index and exchanging the role of form and frame
indices. 
The equations of motion (3.4.5) imply that 
(i) $t_{[b\vert a_{1}a_{2}a_{3}]}= \partial_{[b} C_{a_{1}a_{2}a_{3}]}\,$, thereby 
re-introducing a three form on-shell, and (ii)  the mixed-symmetric part 
${t}_{ a_{1}a_{2}a_{3} \Vert b} = 3\,\partial_{b}\partial_{[a_{1}}\lambda_{a_{2}a_{3}]}\,$. 
But this is precisely in the form of its residual gauge 
transformations in the gauge where $A_{[3]}=0\,$, 
so that ${t}_{a[3]\Vert b} $ is pure gauge and does not carry any local  
degree of freedom. 
\par
\medskip

Let us demonstrate that the above action makes contact with the unfolded formalism given earlier in this 
section. Equation (3.4.6)  can be written as
$$
{\rm d} e_{[9]}^{a[3]} + h_{b_{1}}\wedge\ldots\wedge h_{b_{7}}\,\wedge
\tilde\omega_{[3]}{}^{a[3]b[7]} = 0\;.
\eqno(3.4.11)
$$
where
$$
\tilde\omega_{[3]}{}^{a[10]} = -{1\over 7!}\,h^{a_{1}}h^{a_{2}}h^{a_{3}} F_{c[4]}\,
\epsilon^{a[7]c[4]} 
\eqno(3.4.12)$$
plays the role of the connection appearing in (3.3.2). 
More precisely, it is the part of $\omega_{c[3]}{}^{a[3]}{}_{b[7]}$ that is antisymmetrised in 
its ten indices that are written as the two index blocks  $c[3]$ and $b[7]$ that appears is (3.3.2), 
therefore, we rewrite 
$$
\tilde\omega_{c[3]}{}^{a[3]}{}_{c[7]} = {3!7! \over 10!}\,\epsilon_{c[7]f}\,F^{fa[3]}\;,
\eqno(3.4.13)$$
while performing the antisymmetrisation over the ten indices $a[10]$ 
on the right-hand side of (3.4.12) explicitly gives
$$
\tilde\omega_{c[3]}{}^{a[10]}= {4! \over 10!}\,\epsilon^{a[10]b}\,F_{bc[3]}\;.
\eqno(3.4.14)$$
\par
\medskip

Note that the component of  $e_{b[9]|a[3]}$ that transforms in the tensor product  
$[10]\otimes [2]\,$ of $GL(11)$ is pure gauge on-shell, as can be seen by 
suitably projecting (3.4.6) and using (3.4.10). 
Thus, it is only the $GL(11)\,$-irreducible component $A_{b[9]\Vert a[3]}\,$ of $e$
that is glued to the zero-forms on-shell.  
To repeat, the components $e_{[b_{1}\ldots b_{9}\vert}{}_{b_{10}] a_{1}a_{2}}$ 
are pure gauge on-shell and can therefore be eliminated in a gauge, 
leaving only the component $A_{b[9]\Vert a[3]}\,$ with a 
differential gauge invariance in terms of the sole $GL(11)\,$-irreducible 
component $\xi_{a[8]\Vert b[3]}$:  
$\delta_{\xi} A_{b[9]\Vert a[3]} = \partial_{b}\,\xi_{b[8]\Vert a[3]}\,$. 
In this gauge, the field equations (3.4.6) then reduce to  
$$
\partial_{a}A_{a[9]\Vert}{}^{b[3]} = {4! 7! \over 10!}\; \epsilon_{a[10]c}F^{cb[3]}\;.
\eqno(3.4.15)
$$
which we recognise as Equation (3.2.12) in the gauge where $\Xi^{a[9]\Vert}{}_{b[2]}$ is set to zero. 
Upon acting with $\partial^{b}$ and antisymmetrising over the four $b$ indices, 
we get 
$$
\partial^{b}\partial_{a}A_{a[9]\Vert}{}^{b[3]} = {4! 7! \over 10!}\; \epsilon_{a[10]c}
\; \partial^{c}\partial^{[b_{1}}A^{b_{2}b_{3}b_{4}]}\;,
\eqno(3.4.16)$$
which is nothing but the equation (3.2.1) up to an inessential coefficient. 

\medskip 
{\bf 3.5 Higher dualisations on-shell}
\medskip  
In this section we will dualise the higher level components of
the representation space $\cal T$ given in equation  (3.1.1). 
We consider a generic tensor in the list, say $F_{a[4]\Vert b(n)}\,$,
and dualise it on all $n$ indices $b$ so as to define
$$
G^{(n)}{}_{c[10]\vert\vert \ldots \vert\vert f[10]|\vert}{}_{a[4] } := \epsilon_{c[10]b_{1}} 
\epsilon_{d[10]b_{2}}\ldots\,
\epsilon_{f[10]b_{n}} \,F_{a[4]||}{}^{b(n)}\;.
\eqno(3.5.1)
$$
Using the symmetries of $F_{a[4]\Vert b(n)}\,$ we now show that 
the  $G^{(n)}{}_{c[10]\vert\vert \ldots \vert\vert f[10]|\vert}{}_{a[4] } $ 
is $GL(11)\,$-irreducible.  
Explicitly, we find that 
$$
\epsilon^{c[10]f_{1}}\,G^{(n)}{}_{c[10]\vert\vert \ldots \vert\vert f[10]|\vert}{}_{a[4] } 
= (-10!)\,\delta^{f_{1}}_{b_{1}}\;
\epsilon_{d[10]b_{2}}  \ldots \epsilon_{f_{1}f[9]b_{n}} \,F_{a[4]||}{}^{b(n)}=
$$
$$
= 10!\,\epsilon_{d[10]b_{2}}  \ldots 
\epsilon_{f[9]b_{1}b_{n}}\,F_{a[4]||}{}^{b(n)}\equiv 0\;,
\eqno(3.5.2)
$$ 
and 
$$
\epsilon^{c[10]a_{1}}\,G^{(n)}{}_{c[10]\vert\vert \ldots \vert\vert f[10]|\vert}{}_{a[4] } 
= (-10!)\,\delta^{a_{1}}_{b_{1}}\;
\epsilon_{d[10]b_{2}}  \ldots \epsilon_{f[10]b_{n}} \,F_{a[4]||}{}^{b(n)}=
$$
$$
= (-10!) \,\epsilon_{d[10]b_{2}}  \ldots 
\epsilon_{f[10]b_{n}}\,F_{b_{1}a[3]||}{}^{b_{1}b_{2}\ldots b(n)}\equiv 0\;.
\eqno(3.5.3)
$$ 
We adopt the short-hand notation in which 
$G^{(n)}{}_{c[10]\vert\vert \ldots \vert\vert f[10]|\Vert}{}_{a[4] }$ 
is denoted by $G^{(n)}_{[10,\ldots,10,4]}\,$,
where the number of columns of height 10 is $n$. 
This object satisfies the trace conditions  
$$
{\rm Tr}_{ij}^{10} G^{(n)}_{[10,\ldots,10,4]} = 0\;,\quad 1\leq i < j \leq n\;,
\eqno(3.5.4)
$$
$$
{\rm Tr}_{i \,n+1}^{4} G^{(n)}_{[10,\ldots,10,4]} = 0\;,\quad 1\leq i  \leq n\;,
\eqno(3.5.5)
$$
where we recall that the symbols $Tr^n_{ij}$ means that one takes an $n$  trace between  indices in the $i$ and  $j$th column. To show the first relation we note that 
$$
\delta_{c_{1}}^{f_{1}}\ldots \delta_{c_{p}}^{f_{p}}\, G^{(n)}{}^{c[10]||}{}_{\ldots \vert\vert f[10]|\vert}{}_{a[4] } = 
-p!(11-p)!\; \delta_{c[10-p]b_{1}}^{f[10-p]b_{n}} 
\epsilon_{d[10]b_{2}}\ldots\,\,F_{a[4]||}{}^{b(n)}\;,
\eqno(3.5.6)
$$
where $\delta^{c[p]}_{d[p]} = \delta^{[c_{1}}{}_{d_{1}}\ldots\delta^{c_{p}]}{}_{d_{p}}\,$, 
so that the expression on the left-hand side vanishes only if the 
antisymmetrised product of Kronecker deltas on the right-hand side of the equation contains $\delta^{b_{n}}{}_{b_{n}}\,$, namely only when $p=10\,$.
\par
To show equation  (3.5.5) we note that 
$$
G^{(n) \,a[p]c[10-p]}{}_{||d[10]|| \ldots || f[10]|\vert}{}_{a[4] } = \epsilon^{a[p]c[10-p]b_{1}} 
\epsilon_{d[10]b_{2}}\ldots\,
\epsilon_{f[10]b_{n}} \,F_{a[4]||b_{1}}{}^{b_{2}\ldots b_{n}}\;,
\eqno(3.5.7)
$$
which only gives zero when $p=4\,$, i.e. when all the four indices $a$'s of 
$F_{a[4]||b_{1}}{}^{b_{2}\ldots b_{n}}\,$ are antisymmetrised with one of the $n$ indices in 
the set $b(n)\,$. Obviously, the result is unchanged if one took four traces involving any 
another of the $n$ columns of length 10 in $G^{(n)}_{[10,\ldots,10,4]}\,$. 
\par
We now consider the derivatives acting on 
$G^{(n)}{}_{c[10]\vert\vert \ldots \vert\vert f[10]|\vert}{}_{a[4] } $.  
In particular we observe that 
$$
\partial_{[a_{1}} G^{(n) \,c[10]||d[10]|| \ldots || f[10]}{}_{||a_{2}\ldots a_{5}]} = 
\epsilon^{c[10]b_{1}} \epsilon^{d[10]b_{2}}\ldots\, \epsilon^{f[10]b_{n}} \,
\partial_{[a_{1}}F_{a_{2}\ldots a_{5}]||}{}^{b(n)}\; = 0\;,
\eqno(3.5.8)
$$
which is a consequence of the differential equations (3.1.10) obeyed by the hierarchy of tensors
in the set $\cal T$ in (3.1.1) together with 
the $GL(11)$-irreducible symmetry properties (3.1.3) of these tensors.  
Whereupon using the generalised Poincare lemma [28] 
the  $GL(11)\,$-irreducible tensors $G^{(n)}_{[10,\ldots,10,4]}$ can be expressed as 
generalised curvature tensors of  $GL(11)\,$-irreducible potentials: 
$$
G^{(n)}{}_{c[10]||\ldots ||f[10],a[4]} = 4(10)^{n}\partial_{c}\ldots \partial_{f}\partial_{a}\,
A^{(n)}{}_{c[9]||\ldots ||f[9]||a[3]}\;,
\eqno(3.5.9)
$$
The tensors $G^{(n)}(A^{(n)})$ are invariant under the following gauge transformations
$$
\delta_{\lambda} A^{(n)}_{[9,\ldots,9,3]} = {\rm d}^{\{ n \} } \lambda^{(n)}_{[9,\ldots,9,8,3]} + 
{\rm d}^{\{ n+1 \} } \lambda^{(n+1)}_{[9,\ldots,9,2]}\;.
\eqno(3.5.10)
$$
where 
$$
{\rm d}^{\{ n \} } \lambda^{(n)}_{[9,\ldots,9,8,3]} \rightarrow \;9\partial_{c^{n}_{1}}\Lambda_{c^{1}[9]||\ldots||c^{n-1}[9]||c^{n}[8]||a[3]} + 
9x\,\partial_{c^{n-1}_{1}}\Lambda_{c^{1}[9]||\ldots||c^{n-1}[8]c^{n}_{1}||c^{n}[8]||a[3]}
$$
$$
+\ldots + 9x\,\partial_{c^{1}_{1}}\Lambda_{c^{1}[8]c^{n}_{1}||\ldots||c^{n-1}[9]||c^{n}[8]||a[3]}\;, 
\qquad x = -{9 \over 8}\;.
\eqno(3.5.11)
$$
The first term on the right-hand side of equation (3.4.10) can be depicted by the Young 
tableau 
$$
\vbox{\offinterlineskip\cleartabs
\def\hr{\vrule height .4pt width 2.5em}
\def\vr{\vrule height15pt depth 5pt}
\def\cc#1{\hfill#1\hfill}
\+ \hr&\hr&\hr&\hr&\cr
\+ \vr\cc{${c_1}$}&\vr\cc{$\ldots $}&\vr\cc{${f_1}$}&\vr\cc{${a_1}$}&\vr \cr
\+ \hr&\hr&\hr&\hr&\cr
\+ \vr\cc{$c_2$}&\vr\cc{$\ldots$} &\vr \cc{$d_2$}&\vr\cc{$a_2$}&\vr\cr
\+ \vr\cc{${c_2}$}&\vr\cc{$\ldots $}&\vr\cc{${f_2}$}&\vr\cc{${a_2}$}\vr& \cr
\+\hr&\hr&\hr&\hr\cr
\+ \vr\cc{$\vdots$}&\vr\cc{$\ldots$} &\vr\cc{$\vdots$}&\vr &\cr
\+ \hr&\hr&\hr&&\cr
\+ \vr\cc{$\vdots$}&\vr\cc{$\ldots$} &\vr\cc{$\vdots$}&\vr&\cr
\+ \hr&\hr&\hr&&\cr
\+ \vr\cc{$c_{9}$}&\vr\cc{$\ldots$} &\vr\cc{$c_{9}$}&\vr&\cr
\+ \hr&\hr&\hr& &\cr}
\eqno(3.5.12)$$
Finally, the  curvatures $G^{(n)}$, $n=0,1,\ldots$ are related by
$$
\partial_{g}G^{(n)}{}_{c[10]||\ldots ||f[10]||a[4]} = -{1\over 10!}\epsilon_{g}{}^{b[10]}\,
G^{(n+1)}{}_{b[10]||c[10]||\ldots ||f[10]||a[4]}
\eqno(3.5.13)
$$
leading to a corresponding duality relation for the first differentials of the various 
potentials:
$$
\epsilon_{gb[10]}\,\partial^{g}A^{(n)}{}_{c[9]||\ldots ||f[9]||a[3]} = 10\,\partial_{b} 
A^{(n+1)}{}_{b[9]||c[9]||\ldots ||f[9]||a[3]}  
+ 3 \partial_{a}\Lambda^{(1)}{}_{g[10] , c[9]||\ldots ||f[9]||a[3] }\;
$$
$$
+Y(\partial_{f} \Lambda^{(2)}{}_{g[10] , c[9]||\ldots ||f[8]||a[3]})
\eqno(3.5.14)
$$
where the symbol $Y$ means the projection of the $9n+3$ indices 
$\{c[9]||\ldots ||f[9]||a[3]\}$ on the $GL(11)$ Young tableau with $n$ columns of height $9$ 
and one of height $3\,$. Drawing from the experience we gained from the 
frame-like formulation of the gauge field $A_{[9,3]}\,$, we expect that the 
first-order duality relation (3.5.14) become free of inhomogeneous term 
when expressed in terms of the frame-like frame fields and connections. 

\medskip
\medskip
{\bf {4 Discussion }}
\medskip
In this paper we have shown how the manifestly Lorentz and gauge covariant formulation of the irreducible representations of the Poincar\'e group leads naturally to a description of the dynamics of the massless point particle in terms of an infinite number of gauge fields which obey first order duality relations. 
Gauge fields of this type were  automatically contained  within the $E_{11}$ non-linear realisation which is conjectured to be a symmetry of the underlying theory of strings and branes. The $E_{11}$ symmetry acts on the infinite number of gauge fields rotating them into each other and this part of the symmetry can be thought of an extension of what we usually regard as a duality symmetry.  
\par
Duality symmetries have played an important part in theoretical physics and one may hope that the extension of the symmetry  given in this paper may prove useful in future work. Certainly it will act as a very useful guide when formulating the equations of motion that follow from the $E_{11}$ non-linear realisation.

\medskip
{\bf {Acknowledgement }}
\medskip
PW wishes to thank Paul Cook for discussions and the SFTC for support from Consolidated grant 
number ST/J002798/1. NB is Research Associate of the Fonds de la Recherche 
Scientifique$\,$-FNRS (Belgium). His work was partially 
supported by a contract ``Actions de Recherche concert\'ees$\,$-Communaut\'e fran\c{c}aise 
de Belgique'' AUWB-2010-10/15-UMONS-1.
 NB thanks the Mathematics Department of King's College London for hospitality, 
while PS and PW thank the hospitality of the Service de M\'ecanique et Gravitation of UMONS where 
parts of this work were done.  PS and PW acknowledge the support of the Conicyt grant 
DPI 201 401 15. 

\medskip
{\bf {Appendix}}
\medskip
In this appendix we give some of the notation used in this paper. While these definitions are given in the text it may not always be easy for the reader to find them  and so we collect them here for easy reference. The first few sections of the paper are written without using elaborate notation so that the reader can get used to the subject, but as the paper progresses we need more and more indices and so we introduce a shorthand notation. 
\par
We separate blocks of antisymmetrised or symmetrised indices on the fields by putting a double bar, for example $A_{a_1a_2a_3\Vert b_1\ldots a_9}$. 
We eventually use a shorthand for blocks of antisymmetric and symmetric indices by  denoting  $A_{a[n]}\equiv A_{[a_1\ldots a_n]} \equiv A_{a_1\ldots a_n}$  for  blocks of  antisymmetric  indices and 
$S_{a(n)}\equiv S_{(a_1\ldots a_n)} \equiv S_{a_1\ldots a_n}$ for symmetrised indices. We use the  
strength-one (anti)symmetrisation convention.  
\par
In the early sections of the paper we denote antisymmetrisation in the usual way that is $F_{a_1a_2a_3a_4}=4 
\partial_{[a_1} A_{a_2a_3a_4]}$. However, once we have more indices to cope with we adopt the convention that when  an index with  the same Latin label occurs in the same up, or down,  position in an equation, it is automatically antisymmetrised. For example, when we write 
$F_{a[4]} = 4\,\partial_{a}A_{a[3]}$ we automatically mean $F_{a_1a_2a_3a_4}=4 
\partial_{[a_1} A_{a_2a_3a_4]}$.
\par
When we are discussing forms we label the degree of the form by a number in square brackets written as a subscript, for example $A_{[3]} = {1 \over 3!}\,dx^{a_{1}}\wedge dx^{a_{2}}\wedge dx^{a_{3}}\,A_{a_1a_2a_3}$.

\medskip
{\bf {References}}
\medskip
\item {[1]} P.A.M Dirac, {\it Quantized Singularities in the Electromagnetic Field},
Proc. Roy. Soc, {\bf A133} (1931) 60
\item {[2]} G. 't Hooft,  {\it Magnetic Monopoles in Unified Gauge
Theories},  Nucl. Phys. {\bf B79} (1974) 276;  A. M. Polyakov, {\it
Particle Spectrum in the Quantum Field Theory}, JETP Lett. {\bf 20}
(1974) 194.
\item {[3]} C. Montonen and D. Olive, {\it Magnetic Monopoles as Gauge Particles?}, 
Phys. Lett. {\bf 72 B} (1977) 117.
\item {[4]}E. Cremmer and B. Julia,
{\it The $N=8$ supergravity theory. I. The Lagrangian},
Phys.\ Lett.\ {\bf 80B} (1978) 48.
\item {[5]} M. K. Gaillard and B. Zumino, {\it Duality Rotations for Interacting Fields}, 
  Nucl.\ Phys.\ B {\bf 193} (1981) 221.
\item {[6]} P. West, {\it $E_{11}$ and M Theory}, Class.Quant.Grav.
{\bf 18}  (2001) 4443,  hep-th/0104081.
\item {[7]} T.~Curtright, {\it Generalized Gauge Fields}, Phys.\ Lett.\ B {\bf 165} (1985) 304.
\item{[8]} C.~M.~Hull,  C. Hull, {\it   Strongly Coupled Gravity and Duality}, Nucl.Phys. {\bf B583} (2000) 237, hep-th/0004195. 
\item{[9]}  N.~Boulanger, S.~Cnockaert and M.~Henneaux, {\it A note on spin s duality}, 
JHEP {\bf 0306} (2003) 060 [hep-th/0306023].
\item{[10]} P. West, {\it Introduction to Strings and Branes}, Cambridge University Press 2012, chapter 17. 
\item {[11]} F.  Riccioni and P. West, {\it The $E_{11}$ origin of all
maximal supergravities},  JHEP {\bf 0707} (2007) 063;  arXiv:0705.0752.
\item{[12]} E.~A.~Bergshoeff, I.~De Baetselier and T.~A.~Nutma,
 {\it E(11) and the embedding tensor}, JHEP {\bf 0709} (2007) 047,
arXiv: hep-th/0705.1304.  
\item {[13]} F. Riccioni and P. West,  {\it Dual fields and $E_{11}$},
Phys.Lett. {\bf 645B} (2007) 286-292,  hep-th/0612001.
\item {[15]} F. Riccioni, D. Steele and P. West, 
{\it  Duality Symmetries and $G^{+++}$ Theories}, 

Class.Quant.Grav.25 (2008) 045012, arXiv:0706.3659. 
\item {[14]} N. Boulanger, P. Cook and D. Ponomarev, {\it Off-Shell Hodge Dualities in Linearised Gravity and E11}, JHEP {\bf 1209} (2012) 089 [arXiv:1205.2277 [hep-th]].
\item{[16]} E.~P.~Wigner,
  {\it On Unitary Representations of the Inhomogeneous Lorentz Group}, 
   Annals Math.\  {\bf 40} (1939) 149 [Nucl.\ Phys.\ Proc.\ Suppl.\  {\bf 6} (1989). 
\item {[17]}  V.~Bargmann and E.~P.~Wigner,
  {\it Group Theoretical Discussion of Relativistic Wave Equations}, 
  Proc.\ Nat.\ Acad.\ Sci.\  {\bf 34} (1948) 211.
\item{[18]} X.~Bekaert and N.~Boulanger,
  {\it Mixed symmetry gauge fields in a flat background}, in the Proceedings of 
  the International Seminar on Supersymmetries and Quantum Symmetries SQS 03, 
   24 -- 29 July 2003, Dubna, Russia, [hep-th/0310209]; 
   {\it Tensor gauge fields in arbitrary representations of GL(D,R). II. Quadratic actions},
  Commun.\ Math.\ Phys.\  {\bf 271} (2007) 723 [hep-th/0606198]; 
  {\it The Unitary representations of the Poincar\'e group in any spacetime dimension}, 
  Proceedings of the $2^{nd}$ Modave Summer School on Mathematical Physics (2006), 
  International Solvay Institutes, 50 pp. [hep-th/0611263].
\item{[19]} 
M.~A.~Vasiliev,
  {\it Cubic interactions of bosonic higher spin gauge fields in AdS(5)}, 
  Nucl.\ Phys.\ B {\bf 616} (2001) 106 [Erratum-ibid.\ B {\bf 652} (2003) 407]
  [hep-th/0106200];  
  {\it Actions, charges and off-shell fields in the unfolded dynamics 
approach}, 
Int.\ J.\ Geom.\ Meth.\ Mod.\ Phys.\  {\bf 3} (2006) 37, [hep-th/0504090].
\item{[20]} E.~D.~Skvortsov,
  {\it Mixed-Symmetry Massless Fields in Minkowski space Unfolded}, 
  JHEP {\bf 0807} (2008) 004 [arXiv:0801.2268 [hep-th]].  
\item{[21]} K.~B.~Alkalaev, M.~Grigoriev and I.~Y.~Tipunin,
  {\it Massless Poincar\'e modules and gauge invariant equations},
  Nucl.\ Phys.\ B {\bf 823} (2009) 509 [arXiv:0811.3999 [hep-th]].
\item{[22]} N.~Boulanger, C.~Iazeolla and P.~Sundell,
 {\it Unfolding Mixed-Symmetry Fields in AdS and the BMV Conjecture: I. General 
 Formalism}, JHEP {\bf 0907} (2009) 013 
 
 [arXiv:0812.3615 [hep-th]].
 \item{[23]} W. Siegel and B. Zwiebach, {\it Gauge string fields from the light cone}, 
 Nucl. Phys. {\bf B} 282, 125 (1987); 
W. Siegel, {\it{Fields}},  hep-th/9912205, Chapters II.B and 
XII.A.  
\item{[24]} N.~Boulanger and D.~Ponomarev,
 {\it Frame-like off-shell dualisation for mixed-symmetry gauge fields}, 
 J.\ Phys.\ A {\bf 46} (2013) 214014 [arXiv:1206.2052 [hep-th]]. 
\item {[25]} M. Hamermesh, 
{\it{Group theory and its application to physical problems}},  Dover, New York, (1989).
\item {[26]} W. Fulton and J. Harris, {\it{Representation Theory: A First Course}}, 
Springer (Graduate Texts in Mathematics / Readings in Mathematics); 
Corrected edition (1991).
\item{[27]} X.~Bekaert, N.~Boulanger and D.~Francia, {\it Mixed-symmetry multiplets and higher-spin curvatures},  arXiv:1501.02462 [hep-th].
\item{[28]} X.~Bekaert and N.~Boulanger,
 {\it Tensor gauge fields in arbitrary representations of GL(D,R): Duality and Poincare lemma}, Commun.\ Math.\ Phys.\  {\bf 245} (2004) 27
  [hep-th/0208058].
\item{[29]} P. West, {\it Very Extended $E_8$ and $A_8$ at low levels, Gravity and
Supergravity}, 
\item{[30]} E. Cremmer, B. Julia and J. Scherk, 
{\it Supergravity Theory in Eleven-Dimensions}, Phys. Lett. {\bf 76B} (1978) 409.
Class.Quant.Grav. 20 (2003) 2393-2406, hep-th/0212291.
\item{[31]} H. Nicolai, P. K. Townsend and P. van Nieuwenhuizen, 
{\it Comments On \ Eleven-dimensional Supergravity }, 
Lett. Nuov. Cim.30 (1981) 315; 
 I. Bandos, N. Berkovits and  D. Sorokin,  {\it Duality symmetric eleven-dimensional  supergravity and its coupling to M-branes}, Nucl.Phys. B522 (1998) 214, hep-th/9711055.

\end